\documentclass[aps,prb,twocolumn,superscriptaddress,floatfix]{revtex4-2}
\usepackage[utf8]{inputenc}
\usepackage[english]{babel}
\usepackage{amsmath}
\usepackage{feynmf}
\usepackage{xcolor}
\usepackage{soul}
\usepackage{graphicx}
\usepackage{amsfonts}
\usepackage{blindtext}
\usepackage{appendix}
\usepackage[colorlinks=true,unicode=true,allcolors=blue]{hyperref}
\usepackage{siunitx}
\graphicspath{{figs/}}

\usepackage[normalem]{ulem}

\renewcommand{\vec}[1]{\mathbf{#1}}
\newcommand{\vk}{{\vec k}}

\newcommand{\eps}{{\varepsilon_\vk}}
\newcommand{\E}{{E_\vk}}

\begin{abstract}
    In pump-probe spectroscopies, THz pulses are used to quench a system, which is subsequently probed by either a THz or optical pulse. In contrast, third harmonic generation experiments employ a single multicycle driving pulse and measure the induced third harmonic. In this work, we analyze a new spectroscopy setup where both, a quench and a drive, are applied and 2D spectra as a function of time and quench-drive-delay are recorded. We calculate the time evolution of the nonlinear current generated in the superconductor within an Anderson-pseudospin framework and characterize all experimental signatures using a quasi-equilibrium approach.
    We analyze the superconducting response in Fourier space with respect to both the frequencies corresponding to the real time and the quench-drive delay time. In particular, we show the presence of a transient modulation of higher harmonics, induced by a wave mixing process of the drive with the quench pulse, which probes both quasiparticle and collective excitations of the superconducting condensate. 
\end{abstract}

\begin{document}

\title{Quench-drive spectroscopy and high-harmonic generation in
BCS superconductors} 
\author{Matteo Puviani}
\email{matteo.puviani@mpl.mpg.de}
\affiliation{Max Planck Institute for Solid State Research, 70569 Stuttgart, Germany}
\affiliation{Max Planck Institute for the Science of Light, 91058 Erlangen, Germany}
\author{Rafael Haenel}
\email{r.haenel@fkf.mpg.de}
\affiliation{Max Planck Institute for Solid State Research, 70569 Stuttgart, Germany}
\affiliation{Department of Physics and Astronomy, Quantum Matter Institute, University of British Columbia, Vancouver V6T 1Z4, Canada}
.\author{Dirk Manske}
\email{d.manske@fkf.mpg.de}
\affiliation{Max Planck Institute for Solid State Research, 70569 Stuttgart, Germany}

\date{\today}
\maketitle

\section{Introduction} 

The superconducting state of matter is characterized by a zoo of collective modes: These include, among others, Higgs, Leggett, bi-plasmon, and Bardasis-Schrieffer modes \cite{JLowTempPhys.126.901,annurev.varma2015,PhysRevB.92.064508,PhysRevB.100.174515,NatPhys.4.341,TsujiHiggs,schwarz2020theory,PhysRevB.100.140501,PhysRevB.104.144508,Schwarz2020,Gabriele2021}. The study of these modes is currently being established as a new field of collective mode spectroscopy, in the sense that bosonic excitations of the condensate reveal information about the underlying superconducting ground state and symmetry properties of the condensate itself \cite{Schwarz2020, chu2020phase, schwarz2020theory, poniatowski2021spectroscopic}. The Higgs mode, for instance, can be used as a spectroscopic tool to distinguish between different gap symmetries of unconventional superconductors \cite{Schwarz2020}.

The experimental study of collective superconducting modes poses significant challenges. Due to particle-hole symmetry, they generally cannot couple linearly to electromagnetic fields in the spatially homogeneous limit \cite{PhysRevB.92.064508,kamatani2021optical}. Instead, they are activated in a two-photon Raman-like process \cite{PhysRevB.93.180507,PhysRevB.101.220507}. Thus, the main signature of collective modes consists of a renormalization of the nonlinear susceptibility, which can be probed by nonlinear spectroscopic techniques, such as high-harmonic generation, optical Kerr effect, and nonlinear optical conductivity measurements \cite{PhysRevLett.111.057002, PhysRevLett.120.117001, PhysRevB.96.020505}. 

\begin{figure}[ht]
\centering
\includegraphics[width=8cm]{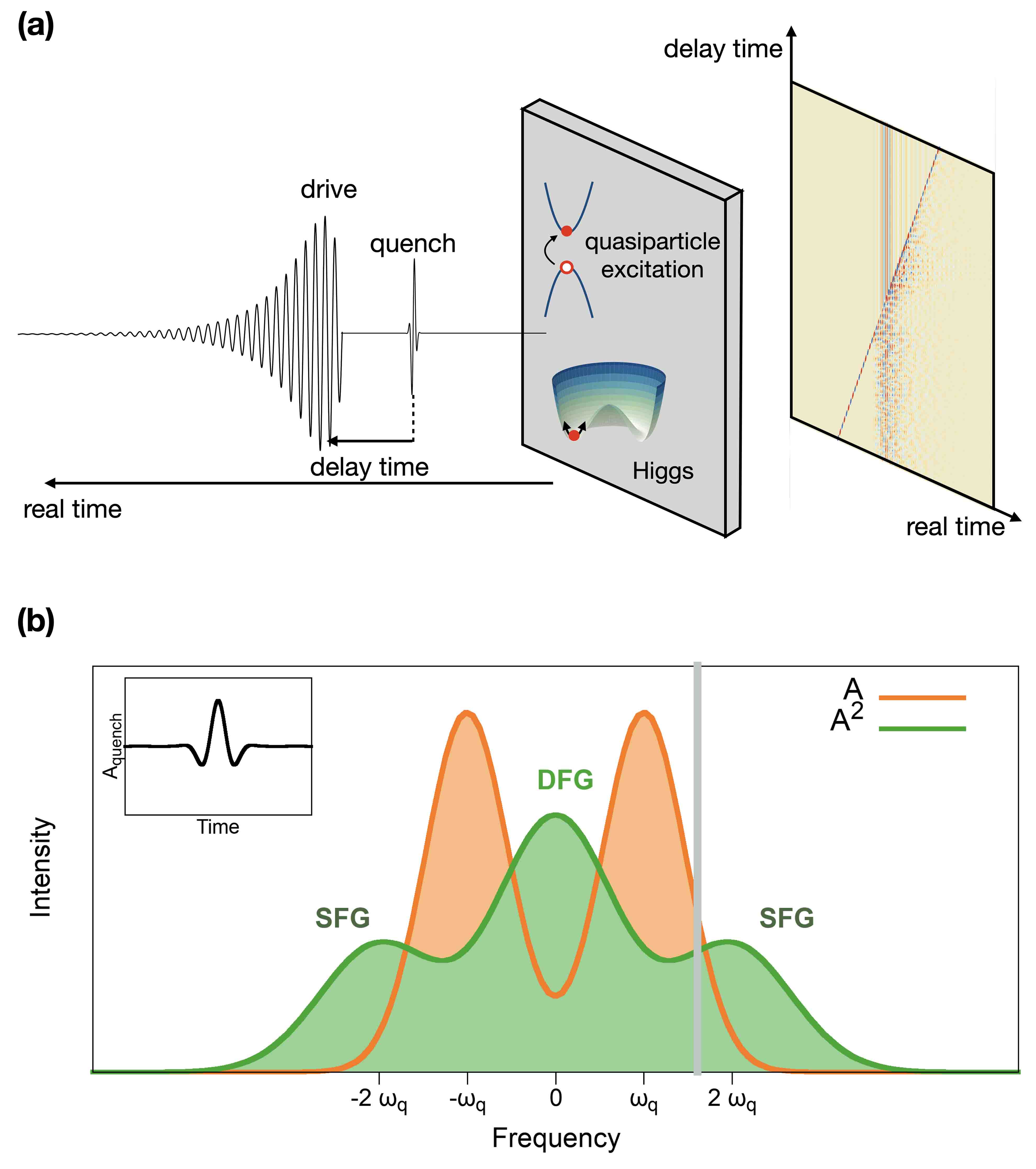}
\caption{\textbf{Quench-drive spectroscopy.} (a) Quench-drive spectroscopy
setup. A single-cylce quench pulse and a multi-cycle drive pulse excite both
Higgs mode and quasiparticles, resulting in a third-harmonic generation (THG) and a dynamical modulation of higher harmonics. In addition, the driving pulse effectively quenches the system, launching Higgs oscillations. In this illustration we show the asymmetric driving pulse. (b) Representation of the frequency spectrum of the quench pulse $A_q (\omega)$ (orange), centered at $\omega = \omega_q$, and the sum (SFG) and difference frequency generated (DFG) pulses $A^2(\omega)$ (green), centered at $\omega = 0$ and $\omega = 2 \omega_q$, respectively. The grey vertical line represents the position of the critical value $2 \Delta$. In the inset the real-time quench pulse is shown.} \label{AA2}
\end{figure}

Generally, two distinct approaches exist to excite collective oscillations of a superconductor: The first is to apply a short-duration quench pulse, $\tau \Delta < 1$ ($\tau$ being the pulse duration and $\Delta$ the energy gap), to suddenly shrink the superconducting gap and excite the system into an out-of-equilibrium state \cite{Schwarz2020}. The second approach uses a single longer pulse, $\tau \Delta \gg 1$, to drive the material into a quasi-steady excited state \cite{PNAS.110.4539, PhysRevLett.120.117001, Schwarz2020}. 

Higher-dimensional spectroscopy techniques have been used to study the nonlinear response of various materials \cite{Cundiff2013,Woerner2013,PhysRevLett.118.207204,Mahmood2021}, but they have rarely been applied to superconductors \cite{PhysRevLett.122.257401}.
Instead, for superconductors, most time-resolved spectroscopies and theoretical studies have focused only on either short quench pulses or a multicycle driving pulse \cite{NatPhys.4.341,PhysRevB.100.165131}, without mixing them simultaneously.
In the present work, we study the full evolution of the nonlinear current in the BCS superconducting state subjected to a spectroscopic setup where \textit{both} a quench and a drive pulse are applied (Fig.~\ref{AA2}(a)).
We show how the spectroscopic data can be clearly analyzed in 2D Fourier space $(\omega,\omega_{\Delta t})$, where the two frequency variables correspond to conjugates of real time $t$ and the pump probe delay $\Delta t$, respectively. While these spectra reduce to the aforementioned pump-probe and THG experiments in certain limits, we argue that quench-drive spectroscopy provides a comprehensive way to experimentally extract information of the nonlinear optical susceptibility and the spectrum of superconducting collective modes.
We stress that one of the experimental advantages of the proposed
quench-drive spectroscopy is that pulse frequencies need not be continuously
scanned across a range of frequencies to probe the system, in contrast to simple
driving protocols \cite{PhysRevResearch.3.L042023}. Instead, to achieve
frequency resolution and collective mode resonance,  only
the time delay between quench and drive pulses needs to be swept.

To solve the equations of motion for a superconductor we employ a pseudospin model, extended to describe the new quench-drive setup. This allows for the simulation of the evolution of the order parameter as well a calculation of the current induced in the superconductor \cite{PhysRevB.92.064508,schwarz2020theory}. 
In addition, we present a diagrammatic approach to systematically interpret the two-dimensional spectra.

The paper is organized as follows: In Section \ref{sec2} we introduce the quench-drive spectroscopy mechanism and explain its features diagrammatically. In Section \ref{theory} we describe the theoretical background. We use the pseudospin model to solve the Heisenberg equation of motion in the presence of an external field for the time-dependent order parameter, and then calculate the generated nonlinear current. In Section \ref{results} we show the numerical results of the nonlinear current for the quench-drive setup in a BCS superconductor. The discussion of results is provided in Section \ref{discussion}. Finally, we give a summary and outlook on future applications and perspectives of quench-drive spectroscopy in Section \ref{conclusion}.

\section{Quench-drive spectroscopy} \label{sec2}
We consider here a clean BCS superconductor without impurity scattering. We are interested in the nonlinear response of the superconductor, 
which is of third order in the external light field in materials with inversion symmetry.
In particular, the quasi-equilibrium third order current is determined by the diamagnetic coupling to light, and its time-dependent expression reads \cite{PhysRevLett.115.157002,schwarz2020theory,PhysRevB.100.165131}
\begin{align}
j(t) = A(t)(\chi_{\rho\rho} * A^2)(t)
\label{eq:nonl-curr}
\end{align}
where $(B * C)(x)=\int dy B(y)C(x-y)$ denotes a convolution integral.
The function $\chi_{\rho\rho}$ is the effective density-density response 
\begin{align}
    \chi_{\rho\rho}(t) = -i\theta(t) \langle [\rho(t),\rho(0)]\rangle
\end{align}
of the operator $\rho = \sum_{\mathbf{k}\sigma} \frac{\partial^2 \epsilon_{\mathbf{k}}}{\partial k_x^2} c_{\mathbf{k}\sigma}^\dagger c_{\mathbf{k}\sigma}$. 
For ease of notation, we have assumed that all applied electromagnetic pulses are polarized along the $\mathbf{x}$-direction, i.e., $\mathbf{A}=A \, \mathbf{\hat{x}}$. In the general case, where $\mathbf{A}$ can have arbitrary polarization, the density-density response becomes a tensor whose structure encodes additional information about material properties. A specific case of cross-polarized pulses is discussed in Apdx.~\ref{Polarization}.

Within the BCS approximation, the gauge-invariant response for a single-band model has been computed in various references \cite{PhysRevB.93.180507,PhysRevB.100.165131,PhysRevLett.115.157002,PhysRevB.92.064508} and a general framework for pump-probe experiments based on a quasi-equilibrium effective action formalism has been developed in Refs.~\cite{PhysRevB.100.165131, NatPhys.4.341}. The density-density response is found to be peaked at the resonance frequency $2\Delta$ of the Higgs mode, where $2\Delta$ is the single-particle superconducting spectral gap. 
It was pointed out, however, that the resonance peak in the density-density response 
is the result of both single-particle contributions, stemming from quasiparticle excitations, and collective mode excitations. Importantly, it was shown that quasiparticles generally give the dominating contribution
to the $2\Delta$-peak in the clean limit, making observation of the Higgs mode difficult \cite{PhysRevB.93.180507}.
Other collective modes of the condensate, such as Leggett \cite{NatPhys.4.341, Benfatto2016Leggett, Haenel2021Froese}, Bardasis-Schrieffer \cite{PhysRevB.104.144508, Schwarz2021Haenel}, or other relative phase modes \cite{poniatowski2021spectroscopic} do contribute significantly to the density-density response and may even persist below the gap. Additionally, the Higgs modes may achieve a sizable signal in the presence of impurities \cite{Silaev2019a,Murotani2019,Seibold2021BenfattoImpurities,Tsuji2020,Haenel2021Froese} or due to additional processes \cite{wang2021transient,PhysRevB.100.104513}.
In the present work, we will not focus on the attribution of spectral weight of the nonlinear response to their various origins and instead discuss spectroscopic measurement of the density-density response $\chi_{\rho\rho}$ as a whole.  

In Fourier space Eq.~\eqref{eq:nonl-curr} can be expressed as
\begin{align}
    j(\omega) &= A * (\chi_{\rho\rho}  (A * A)) 
    = \int \prod_{i=1}^3 d\omega_i \, \delta\big(\omega-\sum_i \omega_i\big)
    \nonumber
    \\
    &\times \,\, A(\omega_1) \chi_{\rho\rho}(\omega_2+\omega_3)
    A(\omega_2)A(\omega_3)
    \label{eqn:cur01}
\end{align}
where the $\delta$-function is a manifestation of energy conservation, i.e., the three photon frequencies $\omega_i$ have to sum up to the frequency $\omega$ of the induced current.
\\
The susceptibility $\chi_{\rho\rho}$ in Eq.~(\ref{eqn:cur01}) enters with a functional dependence on
the frequency variables $\omega_2+\omega_3$. In general, the integration over these
variables scrambles the resonance spectrum of $\chi_{\rho\rho}$ and 
no direct signature of collective modes can be recovered from $j(\omega)$.\\
Two approaches to circumvent this problem exist. First, one may choose
$A$ as a multicycle THz pulse of the form
\begin{align}
	A(t) = A_d(t) = A^0_d \cos(\omega_d t + \phi) \, e^{-t^2/2\tau_d^2}
    \label{eq:pulse_a}
\end{align}
with $\tau_d \omega_d \gg  1$ and $\omega_d\sim\Delta$ such that it has a narrow frequency spectrum of width
$\tau_d^{-1}$ centered around $\pm\omega_d$. Then, the integration variables are
constrained to $\omega_i\approx \pm\omega_d$ and the
susceptibilities are mostly evaluated at $\chi_{\rho\rho}(0)$ and
$\chi_{\rho\rho}(\pm2\omega_d)$ to yield the first or third harmonic,
$j(\pm\omega_d),j(\pm3\omega_d)$.
To map out the functional dependence of $\chi_{\rho\rho}$
one has to vary the driving frequency $\omega_d$. This, however, is not
easily achievable experimentally. Instead, most current experiments 
fix the driving frequency $\omega_d$ and instead attempt to shift the resonance energies contained in $\chi_{\rho\rho}$. For a superconducting mode, this is simply achieved by varying the temperature in the window $(0,T_c)$.
The clear disadvantages of this method are that (1) knowledge of the temperature dependence of the resonances of $\chi_{\rho\rho}$ is required, (2) only modes above $2\omega_d$ are visible, and (3) thermal broadening effects are
substantial.

\begin{figure*}[ht]
\centering
\includegraphics[width=\textwidth]{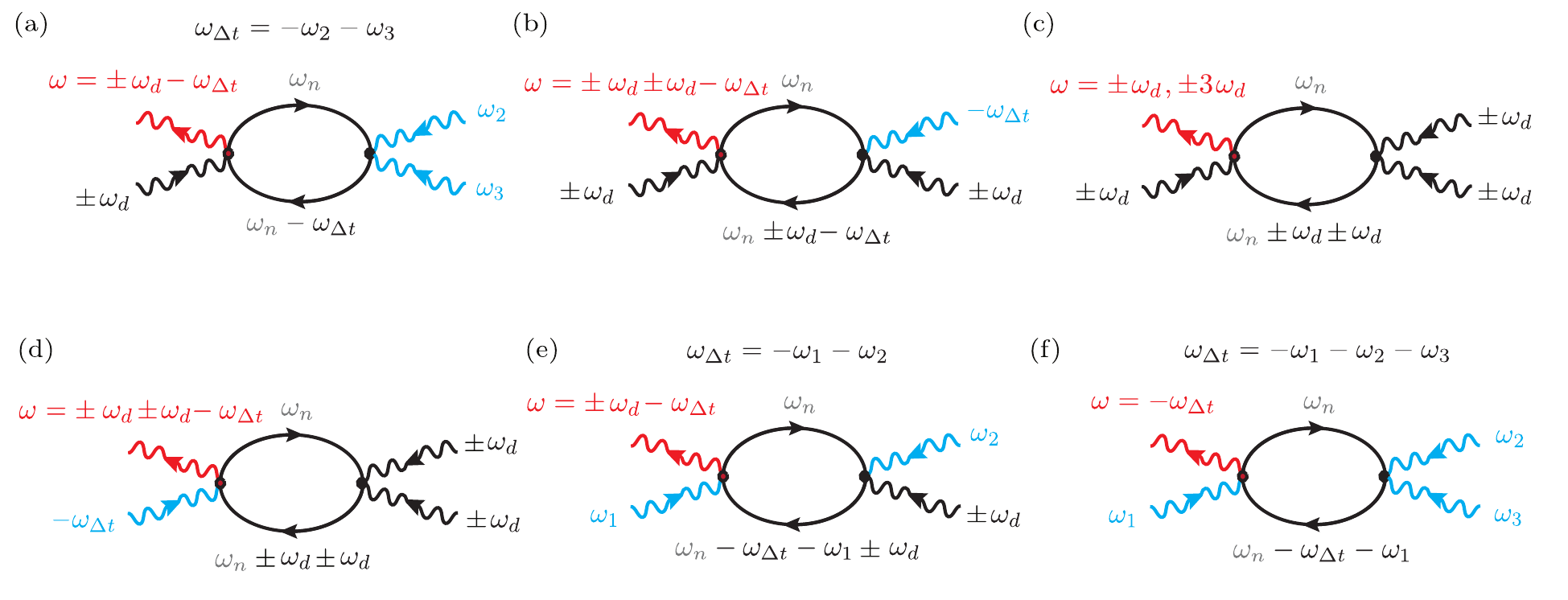}
\caption{\textbf{Diagrammatic representation of the nonlinear processes.} The
wiggly lines represent photons: the black ones correspond from the driving field
$A_d$, the
blue lines correspond to the quench pulse $A_q$, the red line represents the generated current. 
Solid lines denote fermionic bubbles that represent the nonlinear susceptibility
$\chi_{\rho\rho}$. Due to the $\delta$-function constraint in
Eq.~\eqref{eqn:final_exp}, all blue quench frequencies have to add up to
$-\omega_{\Delta t}$. Energy conservation demands that all incoming frequencies
add up to the frequency $\omega$ of the induced current.
}
\label{Diag}
\end{figure*}

The second approach consists of pump-probe setup. Here, we consider a novel pump-probe setup where in addition to a broadband quench pulse $A_q$, the multi-cycle drive pulse $A_d$ is utilized. The quench $A_q$ has the same form as Eq.~\eqref{eq:pulse_a}, but with $\tau_q \ll 1/\Delta$. 
Further details of the pulse shapes are given in Appendix~\ref{pulsesDetails}.
The two pulses are delayed with respect to each other by $\Delta t$ yielding the total field $A(t)=A_q(t+\Delta t) + A_d(t)$ (see Fig.~\ref{AA2} and Fig.~\ref{pulses}). In frequency space, this results in a phase shift,
\begin{align}
    A(\omega) &= e^{i\omega \Delta t} A_q(\omega) + A_d(\omega)
    \nonumber
    \\
    &= \sum_{\alpha=q,d}^{} e^{i \delta_{\alpha,q}\omega \Delta t}
    A_{\alpha}(\omega)
    \label{eqn:pulses2}
\end{align}
where we have introduced the notation $\delta_{\alpha,q}=1$ for $\alpha=q$ and zero
for $\alpha=d$. \\
In a nonlinear THz experiment, the current $j(t)$ is electro-optically sampled as a function of $t$ and Fourier
transformed numerically to obtain $j(\omega)$. Multiple such
traces are recorded for varied $\Delta t$ to assemble the 2D spectrum
$j(\omega;\Delta t)$.
Inserting Eq.~\eqref{eqn:pulses2} into Eq.~\eqref{eqn:cur01}, we obtain
\begin{align}
	j(\omega;\Delta t) &= \int
		\prod_{i=1}^3 d\omega_i \,
	\delta\big(\omega-\sum_{i}\omega_i\big)
		\nonumber
	\\
	& \,\, \times
		\sum_{\alpha_1,\alpha_2,\alpha_3=q,d} 
	\exp \left( 
	i\Delta t\sum_{i=1}^{3}\delta_{\alpha_{i},q}\omega_i
	\right)
	\nonumber
	\\
	& \,\, \times
	A_{\alpha_1}(\omega_1)
	\chi_{\rho\rho}(\omega_2+\omega_3)
	A_{\alpha_2}(\omega_2)
	A_{\alpha_3}(\omega_3) \,,
\end{align}
where we are summing over all combinations $\{\alpha_i\}$ of quench and drive pulse, $A_q$ and
$A_d$, respectively.
We perform a Fourier transform in the parametric time delay
$\Delta t$ to obtain
\begin{align}
	j(\omega; \omega_{\Delta t}) &= \int  
	\prod_{i=1}^3 d\omega_i \,
	\delta\big(\omega-\sum_{i}\omega_i\big)
	\nonumber
	\\
	& \,\, \times
	\sum_{\alpha_1,\alpha_2,\alpha_3=q,d} 
	\delta \big( 
	\omega_{\Delta t}+\sum_{i=1}^3\delta_{\alpha_{i},q}\omega_i
	\big)
	\nonumber
	\\
	& \,\, \times
	A_{\alpha_1}(\omega_1)
	\chi_{\rho\rho}(\omega_2+\omega_3)
	A_{\alpha_2}(\omega_2)
	A_{\alpha_3}(\omega_3) 
	\label{eqn:final_exp}
\end{align}

We can represent the various terms in the sum over $\{\alpha_i\}$
diagrammatically as depicted in Fig.~\ref{Diag}. Here, the current is
represented by a red wiggly line, and quench and probe pulses are depicted by blue and
black wiggly photon lines, respectively. The density-density susceptibility
$\chi_{\rho\rho}$ is represented by a
fermionic bubble whose internal frequency we have labeled $\omega_n$. 

	All external photon lines carry frequencies with profiles
determined by the experimental bandwidth of the pulses $A_{\alpha_i}(\omega_i)$,
where the directionality is marked by arrows of the photon lines. The drive
pulse will constrain the external frequencies to $\sim \pm \omega_d$. 
In fact, we will be assuming a sufficiently narrow-band drive pulse, $\tau_d \gg
1/\Delta$, such that we can approximate 
\begin{align}
	A_d(\omega_i) \sim \delta(\omega_i - \omega_d) + \delta(\omega_i +
	\omega_d) \,.
	\label{eq:apdr}
\end{align}

The key advantage of the pump-probe geometry lies in the 
appearance of the second $\delta$-function
in
Eq.~\eqref{eqn:final_exp} that introduces the 
experimentally accessible variable $\omega_{\Delta t}$. 
We can see this as follows.
When $\alpha_1=d$ and $\alpha_2=\alpha_3=q$, the $\delta$-function presents the
constraint
$\delta(\omega_{\Delta t} + \omega_2 + \omega_3)$
and the density-density correlation function is evaluated
at $\chi(\omega_2+\omega_3) = \chi(-\omega_{\Delta t})$.
Thus, $\chi$ can be pulled out of the integral in Eq.~\eqref{eqn:final_exp} and
the measured current is directly proportional to the $\chi_{\rho\rho}$-response,
whose frequency dependence can be mapped out by sweeping
$\omega_{\Delta t}$.
The diagrammatic representation of this process
is shown in Fig.~\ref{Diag}(a). Making use of approximation~\eqref{eq:apdr}, it
follows from energy conservation that the current is non-zero only along the
lines $\omega = \pm\omega_d - \omega_{\Delta t}$ in 2D frequency space
$(\omega,\omega_{\Delta t})$, where $j$ is given by
\begin{align}
	j(\pm\omega_d-\omega_{\Delta t}, \omega_{\Delta t})
	&\propto
	\chi_{\rho\rho}(-\omega_{\Delta t}) \,.
\end{align}

As similar discussion applies to the process depicted in
Fig.~\ref{Diag}(b). Here, $\alpha_1 = \alpha_3 = d$ and $\alpha_2 = q$.
The susceptibility is evaluated as $\chi(\pm\omega_d - \omega_{\Delta t})$
and determined the current along the lines $\omega=\pm 2 \omega_d -
\omega_{\Delta t}$ and $\omega = -\omega_{\Delta t}$. Explicitly, the current is,
\begin{align}
	&j(\pm2\omega_d-\omega_{\Delta t}, \omega_{\Delta t})
	\propto
	\chi_{\rho\rho}(\pm\omega_d-\omega_{\Delta t})
	\nonumber
	\\
	&j(-\omega_{\Delta t}, \omega_{\Delta t})
	\propto
	\chi_{\rho\rho}(\omega_d-\omega_{\Delta t})
	+
	\chi_{\rho\rho}(-\omega_d-\omega_{\Delta t}) \,.
\end{align}

Figure~\ref{Diag}(c) describes the usual THG process that is independent of the
quench. In 2D Fourier space it yields a signal at $\omega_{\Delta t} = 0$ at the
first and third harmonic frequencies of the drive, $\omega=\pm \omega_d, \pm
3\omega_d$, where the density-density
susceptibility is evaluated at fixed values $0,\pm 2\omega_d$.

The remaining diagrams of Fig.~\ref{Diag} can be separated into two classes. In
Fig.~\ref{Diag}(d) $\chi_{\rho\rho}$ only depends on $\omega_d$ and the discussion of
the THG case applies. 
In Fig.~\ref{Diag}(e-f) the
dependence of $\chi_{\rho\rho}$ on integration variables $\omega_i$ cannot be
removed. As a consequence, the resonance structure of $\chi_{\rho\rho}$ is
washed out by integration and one obtains a mostly constant signal for frequencies much smaller
than the bandwidth of the quench pulse.

In summary, we have shown that the signal $j(\omega,\omega_{\Delta t})$  
falls onto diagonal lines $\omega=\pm n \omega_d -\omega_{\Delta t}$ with $n\in
\{0,1,2\}$ in 2D Fourier space.
From the 2D spectra, one can extract the density-density response according
to
\begin{align}
	j(\pm2m \,\omega_d-\omega_{\Delta t}, \omega_{\Delta t})
	&\propto
	\chi_{\rho\rho}(\pm\omega_d-\omega_{\Delta t}) + c_1
	\nonumber \,,
	\\
	j(\pm\omega_d-\omega_{\Delta t}, \omega_{\Delta t})
	&\propto
	\chi_{\rho\rho}(-\omega_{\Delta t}) + c_2 \,,
	\label{eqn:diagonal-lines-current}
\end{align}
where $m=\{0,1\}$ and the $c_i$ denote the background signal that is mostly constant
in the limit of a broadband quench pulse.

\section{Microscopical description} \label{theory}
Having discussed the phenomenological structure of the nonlinear response in a
quench-drive experiment, let us now microscopically 
investigate the response current of a conventional clean superconductor subject to quench and drive pulses. The solution is obtained solving the Bloch equations derived from the pseudospin model of the BCS Hamiltonian.

\subsection{Equations of motion with quench-drive pulses}
We write the BCS Hamiltonian using the pseudospin formalism \cite{PhysRev.112.1900,PhysRev.117.648,PhysRevB.92.064508,schwarz2020theory} as
\begin{align}
\hat{H} = \sum_\vk \mathbf{b}_\vk \cdot \hat{\mathbf{\sigma}}_\vk \,,
\end{align}
with the pseudospin vector
\begin{align}
\hat{\mathbf{\sigma}}_\vk = \frac{1}{2} \hat{\Psi}_\vk^\dagger \mathbf{\tau} \hat{\Psi}_\vk \,,
\end{align}
which is defined in Nambu-Gor'kov space, with spinor $\hat{\Psi}_\vk = (\hat{c}_{\vk, \uparrow}^\dagger \quad \hat{c}_{- \vk, \downarrow})$ and the Pauli matrices $\mathbf{\tau} = (\tau_1, \tau_2, \tau_3)$. The pseudo-magnetic field is defined by the vector
\begin{align}
\mathbf{b}_\vk = (- \Delta' f_\vk, - \Delta'' f_\vk,  \eps) \,,
\end{align}
where $\eps = \xi_\vk - \mu$, $\xi_\vk$ being the fermionic band dispersion, $\mu$ the chemical potential. The superconducting order parameter $\Delta_\vk = \Delta f_\vk = (\Delta' + i \Delta'') f_\vk$ satisfies the gap equation
\begin{align}
\Delta_\vk = \Delta f_\vk = V f_\vk \ \sum_{\vk'} f_{\vk'} \langle \hat{c}_{-\vk', \downarrow} \hat{c}_{\vk', \uparrow} \rangle \,.
\end{align}
Here, V is the pairing strength, and $f_\vk$ the form factor of the
superconducting order parameter. For s-wave pairing one has $f_\vk =1$.\\
In the presence of an external field represented by the vector potential $\mathbf{A}(t)$, the pseudospin changes in time according to
\begin{align}
\mathbf{\sigma}_\vk (t) = \mathbf{\sigma}_\vk (0) + \delta \mathbf{\sigma}_\vk (t) \,, 
\end{align}
with $\sigma_\vk = \langle \hat{\sigma}_\vk \rangle$ and $\delta \mathbf{\sigma}_\vk (t) = (x_\vk (t), y_\vk (t),  z_\vk (t))$. The external electromagnetic field is included in the pseudo-magnetic field by means of the minimal substitution $\vk \rightarrow \vk - e \mathbf{A}(t)$ in the fermionic energy, resulting in
\begin{align}
\mathbf{b}_\vk (t) = (- \Delta'(t) f_\vk,  - \Delta''(t) f_\vk,  (\varepsilon_{\vk - e \mathbf{A}(t)} + \varepsilon_{\vk + e \mathbf{A}(t)})/2) \,.
\end{align}
The Heisenberg equation of motion for the pseudospin can be written in Bloch form, $\partial_t \mathbf{\sigma}_{\vk} = 2 \textbf{b}_{\textbf{k}} \times \mathbf{\sigma}_{\textbf{k}}$, providing the set of differential equations
\begin{eqnarray}
    \begin{array}{ll}
    \begin{split}
        \partial_t x(t) &= - (\varepsilon_{\vk - e \mathbf{A}} + \varepsilon_{\vk + e \mathbf{A}}) y(t) - \dfrac{f_\vk}{\E} \eps \delta \Delta'' (t)  \\
        &+ 2 \delta \Delta'' (t) f_\vk z(t) \,, \\
        \partial_t y(t) &= 2 \varepsilon_{\textbf{k}} x(t) + 2 ( \Delta + \delta \Delta' (t) ) f_\vk z(t) \\
        &- \ \delta \Delta' (t) f_\vk \dfrac{\eps}{\E} + \dfrac{\Delta f_\vk}{2 \E} (\varepsilon_{\vk - e \mathbf{A}} + \varepsilon_{\vk + e \mathbf{A}} - 2 \eps) \,, \\
        \partial_t z(t) &= -2 \ \Delta f_\vk \ y(t) - \dfrac{\Delta f_\vk^2}{\E} \delta \Delta'' (t) - 2 \delta \Delta'' (t) f_\vk x(t) \,,
    \end{split}
    \end{array}
    \label{Blocheqs}
\end{eqnarray}
where $\delta \Delta (t) = \delta \Delta'(t) + i \ \delta \Delta''(t)$ is the time-dependent variation of the order parameter induced by the external field, such that $\Delta (t) = \Delta + \delta \Delta (t)$.
Here, we assumed a real order parameter at the initial time $t=0$, so that $y(0) = 0$. 
The solution of Eq.~\eqref{Blocheqs} provides the time-dependent evolution of
pseudospins, from which the time-dependent order parameter $\Delta (t)$ and the
generated current $j (t)$ can be calculated. A detailed derivation of the
equations of motion is given in Appendix~\ref{apdx:pseudospin}.

\subsection{Nonlinear current}
The current generated by the superconductor in this quench-drive setup is given by the general expression
\begin{align}
\mathbf{j}(t, \Delta t) = e \sum_{\vk} \mathbf{v}_{\vk - e \mathbf{A}(t, \Delta t)} \langle \hat{n}_\vk (t, \Delta t) \rangle \,,
\end{align}
where the velocity is calculated as $\mathbf{v}_{\vk - e \mathbf{A}(t, \Delta
t)} = \nabla_\vk \varepsilon_{\vk - e \mathbf{A}(t, \Delta t)}$, and the charge
density is defined as $ \langle \hat{n}_\vk (t, \Delta t) \rangle = \langle
\hat{c}^\dagger_{\vk, \uparrow} \hat{c}_{\vk, \uparrow} +  \hat{c}^\dagger_{\vk,
\downarrow} \hat{c}_{\vk, \downarrow}  \rangle (t, \Delta t)$. We can expand the
velocity as a function of the vector potential $\mathbf{A}(t, \Delta t)$, and
expand the current in powers of the external field. 
The first non-vanishing term of the nonlinear current $j^{NL} (t, \Delta t)$ generated by the driving pulse is the third order component
\begin{align}
\mathbf{j}^{(3)} (t, \Delta t) = - 2 e^2 \sum_\vk \sum_{i=x,y} \mathbf{A} (t, \Delta t) \cdot \mathbf{r}_i \ \partial_{k_i} \mathbf{v}_\vk \ z_\vk (t, \Delta t) \,, \label{NLcurrent}
\end{align}
where $z_\vk (t, \Delta t)$ is the third component of the pseudospin vector
$\mathbf{\sigma}_\vk (t, \Delta t)$, that contains the information of the state of the system perturbed by the quench pulse. The unit vector $\mathbf{r}_i$, $i =x,y$, represents the two directions along which the output current is measured.

\section{Numerical results} \label{results}

\begin{figure}[tb]
\centering
\includegraphics[width=8cm]{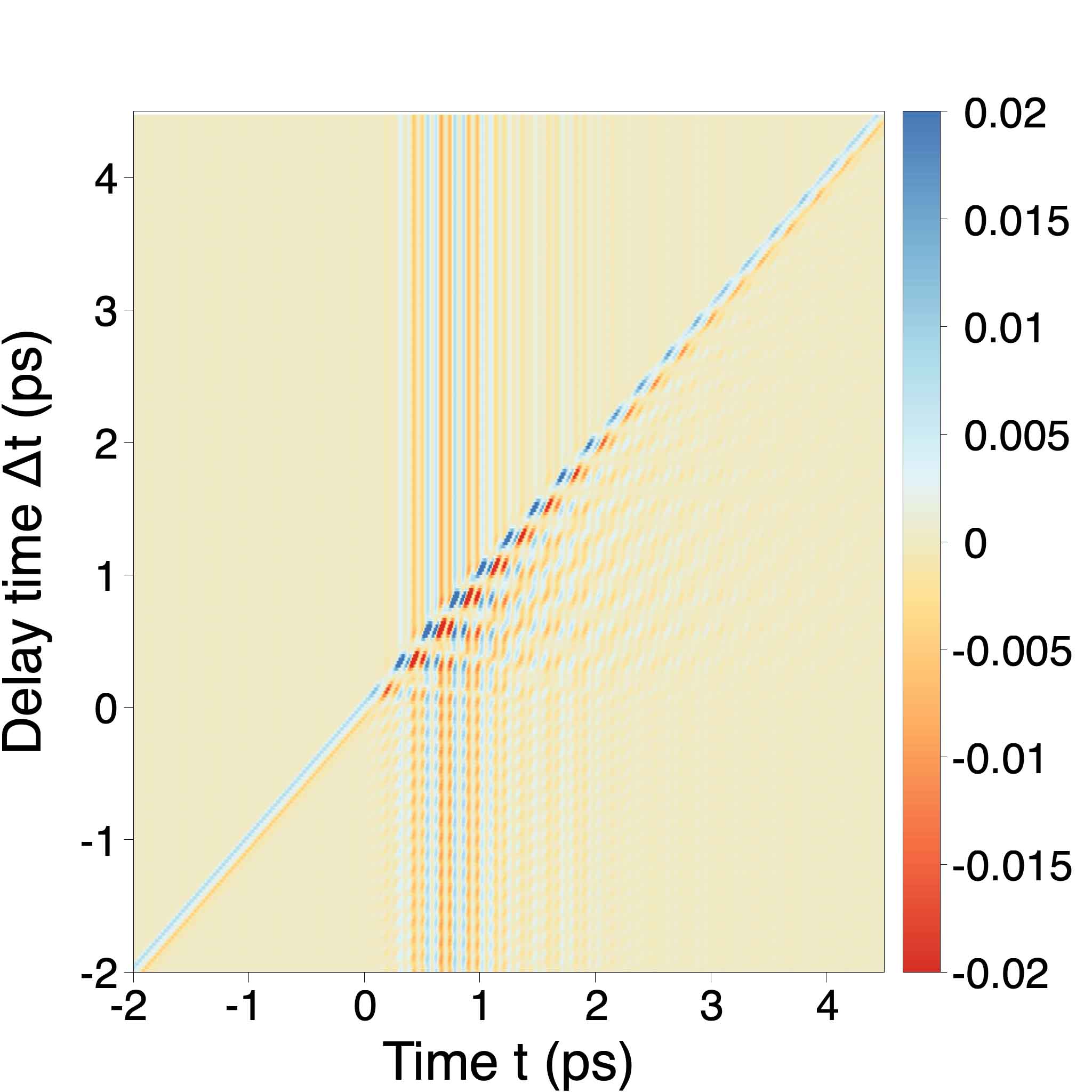}
\caption{\textbf{Time-delay two-dimensional plot of the generated nonlinear current.} 2D plot of the nonlinear output current $j^{NL}(t, \Delta t)$ generated by the driven superconductor, as a function of real time $t$ and the quench-drive delay time $\Delta t$. The narrow diagonal stripes are generated by the quench, while the vertical ones are the response to the drive pulse. The intersection provides a wave mixing pattern for $t, \Delta t \in [0,2]$ ps. Here we used the asymmetric drive pulse, with quench and drive pulse frequencies respectively $\omega_q = 4.77$ THz and $\omega_d = 4.3$ THz.} \label{2Dtt}
\end{figure}

\begin{figure}[tb]
\centering
\includegraphics[width=8cm]{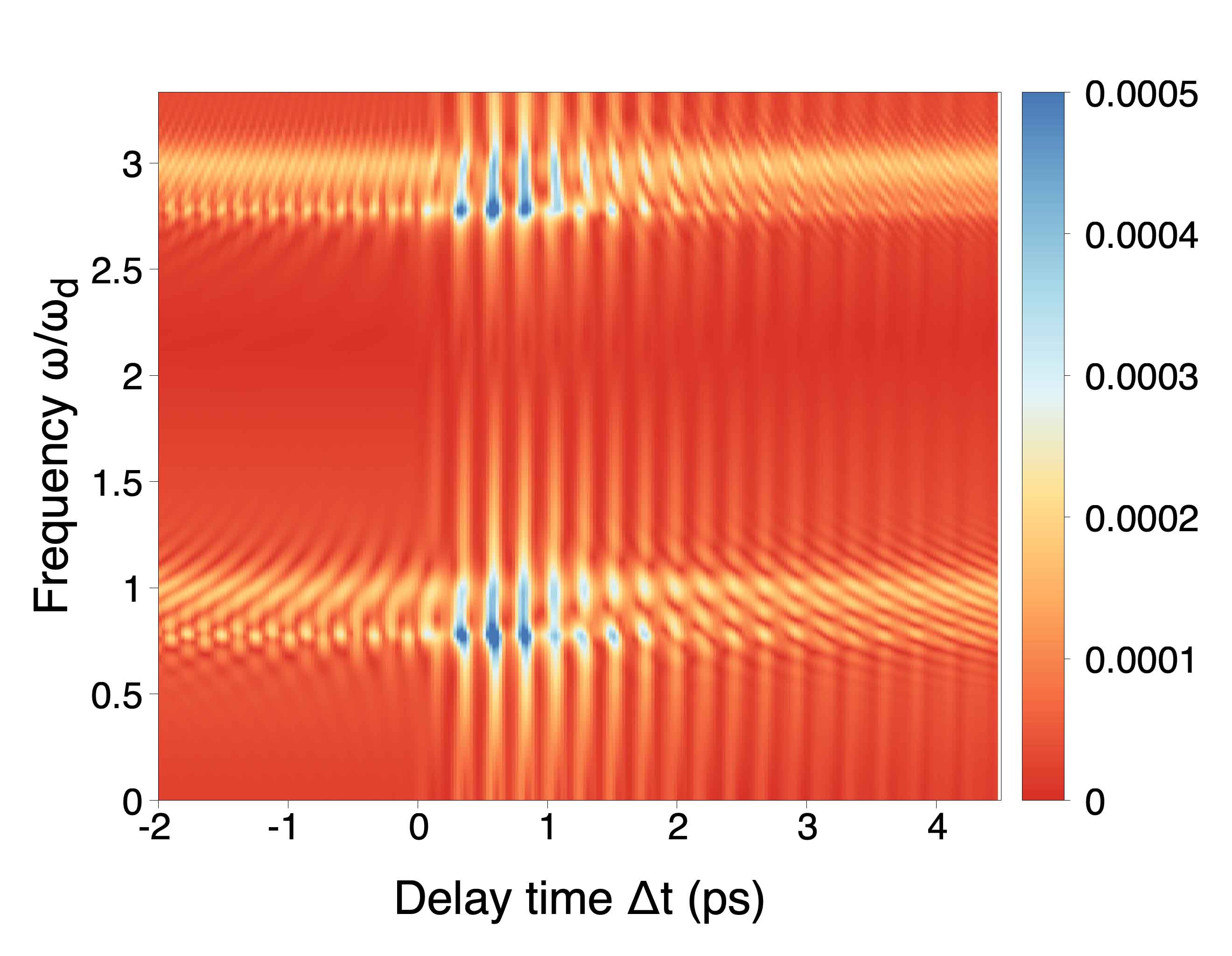}
\caption{\textbf{Frequency-time delay two-dimensional plot of the generated nonlinear current.} 2D plot of the nonlinear current $j^{NL}(\omega, \Delta t)$ generated by the driven superconductor, as a function of relative frequency $\omega / \omega_d$ (with driving frequency $\omega_d = 4.3$ THz) and the quench-drive delay time $\Delta t$. This corresponds to a Fourier transform in the horizontal direction in Fig.~\ref{2Dtt}. The fundamental ($\omega$ = $\omega_d$ = 4.3 THz) and the third harmonic ($\omega$ = $3 \omega_d$ = 12.9 THz) are both modulated in the delay time $\Delta t$.} \label{FFT1-2D}
\end{figure}

We now present the results obtained from the numerical implementation of
time-dependent Bloch equations described in the previous section, solved by
means of a Runge-Kutta-4 algorithm without linearization or further analytical
approximations. 
We used the fermionic band dispersion $\eps = -2 t (\cos{k_x} + \cos{k_y})$ at half-filling, setting the lattice constant $a=1$.
The point $\mu=0$ is special in the sense that it has perfect particle-hole
symmetry as well as a van Hove singularity at the Fermi level. But from the
viewpoint of collective modes, this symmetry point does not bear any special
significance other than presenting a local minimum of the Higgs contribution
compared to single-particle excitations in the nonlinear response
\cite{PhysRevB.93.180507}.

We used the values of $t = 125$ meV for the nearest-neighbor hopping energy,
s-wave order parameter $\Delta_{0} = 15.8$ meV (corresponding to a frequency of
$3.82$ THz), and a summation over the full Brillouin zone with a square sampling
and a total number of $N_\vk = 10^6$ points. For the time-dependent evolution we
used a time-step of $\delta t = 3 \cdot 10^{-4}$ ps, and for the quench-drive
delay $\delta \Delta t = 2.5 \cdot 10^{-2}$ ps. For the quench we used a few
cycles pulse with central frequency $\omega_q = 4.77$ THz, while for the driving
an asymmetric pulse with central frequency $\omega_d = 4.3$ THz, so that both
satisfy the condition $\omega_{q(d)} < 2 \Delta$. Both pulses were considered
linearly polarized along the $\mathbf{x}$-direction, while the maximum amplitude
of the electric field of quench and drive pulse was taken $E_q = 10.5$ kV/cm and
$E_d = 4.7$ kV/cm, respectively, assuming a value for the lattice constant of $a
= 3$ \AA. For more details on the pulses used, see Appendix~\ref{pulsesDetails}. In Appendix~\ref{Polarization} we analyze the 2D spectra for the case of cross polarization of the two pulses, while in Appendix~\ref{AppGauss} we show additional results that were computed using a symmetric Gaussian driving pulse instead of an asymmetric one. 

\subsection{Two-dimensional quench-drive spectroscopy}

\begin{figure}[tb]
\centering
\includegraphics[width=\columnwidth]{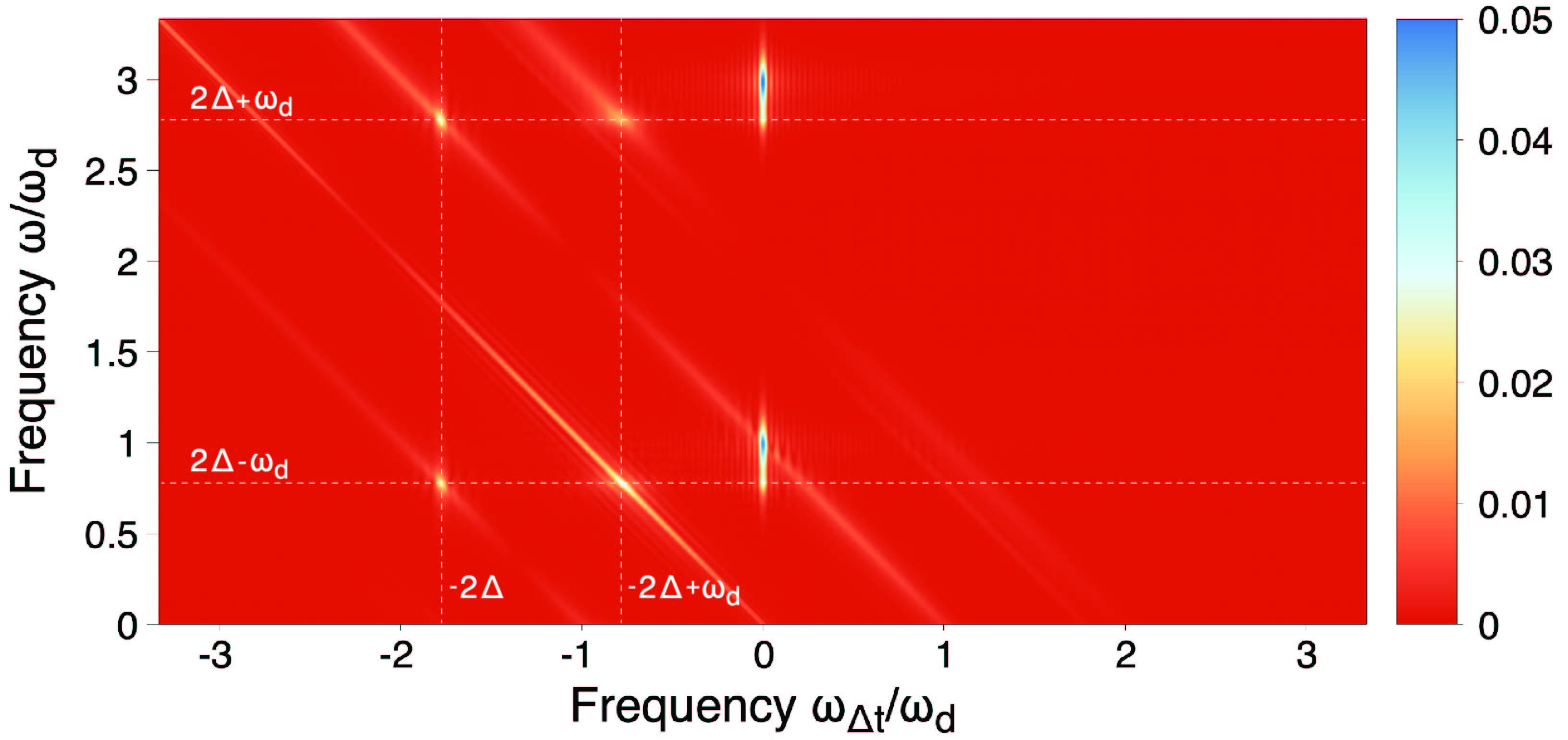}
\caption{\textbf{Two-dimensional Fourier-transformed plot of the nonlinear current.} 2D plot of the generated nonlinear output current intensity $j^{NL}(\omega, \omega_{\Delta t})$ as a function of the real frequency $\omega$ and the quench-drive delay frequency $\omega_{\Delta t}$. It corresponds to the two-dimensional Fourier transform of the data in Fig.~\ref{2Dtt}. The vertical response at $\omega_{\Delta t}=0$ corresponds to the quench-free superconducting signal, namely the high-harmonic generation due to the driving field. The diagonal lines, instead, represent the transient modulation of the higher-harmonics due to the quench-drive wave mixing.} \label{FFT2D}
\end{figure}

We solved the time-dependent equation of motion in the quench-drive spectroscopy setup using the pseudospin model and calculated the nonlinear current generated by the condensate as described in the previous section. The result plotted as a function of the real time evolution $t$ and the quench-drive delay time $\Delta t$, which is measured as the interval between the maximal peaks of the envelopes of the two pulses (see Appendix \ref{pulsesDetails} for more details on the pulse shapes), is shown in Fig.~\ref{2Dtt}. 
We notice that the diagonal line signifies the arrival of the quench pulse,
which overlaps with the drive for times $t = \Delta t$ in the range $t, \Delta t \in [0,2]$ ps. In this region the response is modulated as a function of the delay time $\Delta t$. \\
Next, we Fourier transform the real time variable $t$ into the frequency $\omega$. 
In Fig.~\ref{FFT1-2D} we show the nonlinear generated current intensity as a
function of the quench-drive delay time $\Delta t$, namely $|j^{NL} (\omega,
\Delta t)|$. We notice that both the first and the third harmonic of the
fundamental driving frequency $\omega_d$ are modulated in delay time $\Delta t$, 
with maximum intensity in the interval $0$ ps $\leq \Delta t \leq 2$ ps, 
which corresponds to the range of interference between the quench and the drive pulses,
as shown in Fig. \ref{2Dtt}. Additionally, the signal intensity does not vanish away from $\omega_d$ and $3\omega_d$, where we instead observe a striped pattern, 
with each intensity line tilted towards the central time $\Delta t = 1$ ps. 
\\
These features can be more readily interpreted by plotting the 2D Fourier 
transform of the current, i.e., as a function of the frequency $\omega$ and the delay-time 
frequency $\omega_{\Delta t}$, respectively, shown in Fig.~\ref{FFT2D}. 
In particular, we notice the first and third harmonic as strong peaks in the 
central vertical line, at $\omega_{\Delta t} = 0$, which corresponds to the equilibrium 
response of the driving pulse in absence of the quench. Note that it is sufficient to plot two 
quadrants of the nonlinear current in 2D frequency space, since it follows from 
$j(t,\Delta t) \in \mathbb{R}$ that $j(\omega,\omega_{\Delta t})=j(-\omega,-\omega_{\Delta t})$.
\\
The modulations in $\Delta t$ appear here as broad diagonal lines in 2D frequency 
space as expected from Eq.~\eqref{eqn:diagonal-lines-current}. 
These features correspond to a dynamically generated four-wave mixing 
signal due to both the quench and the drive pulses.

\subsection{High-harmonic generation and transient excitation of the superconductor}

\begin{figure}[tb]
\centering
\includegraphics[width=8cm]{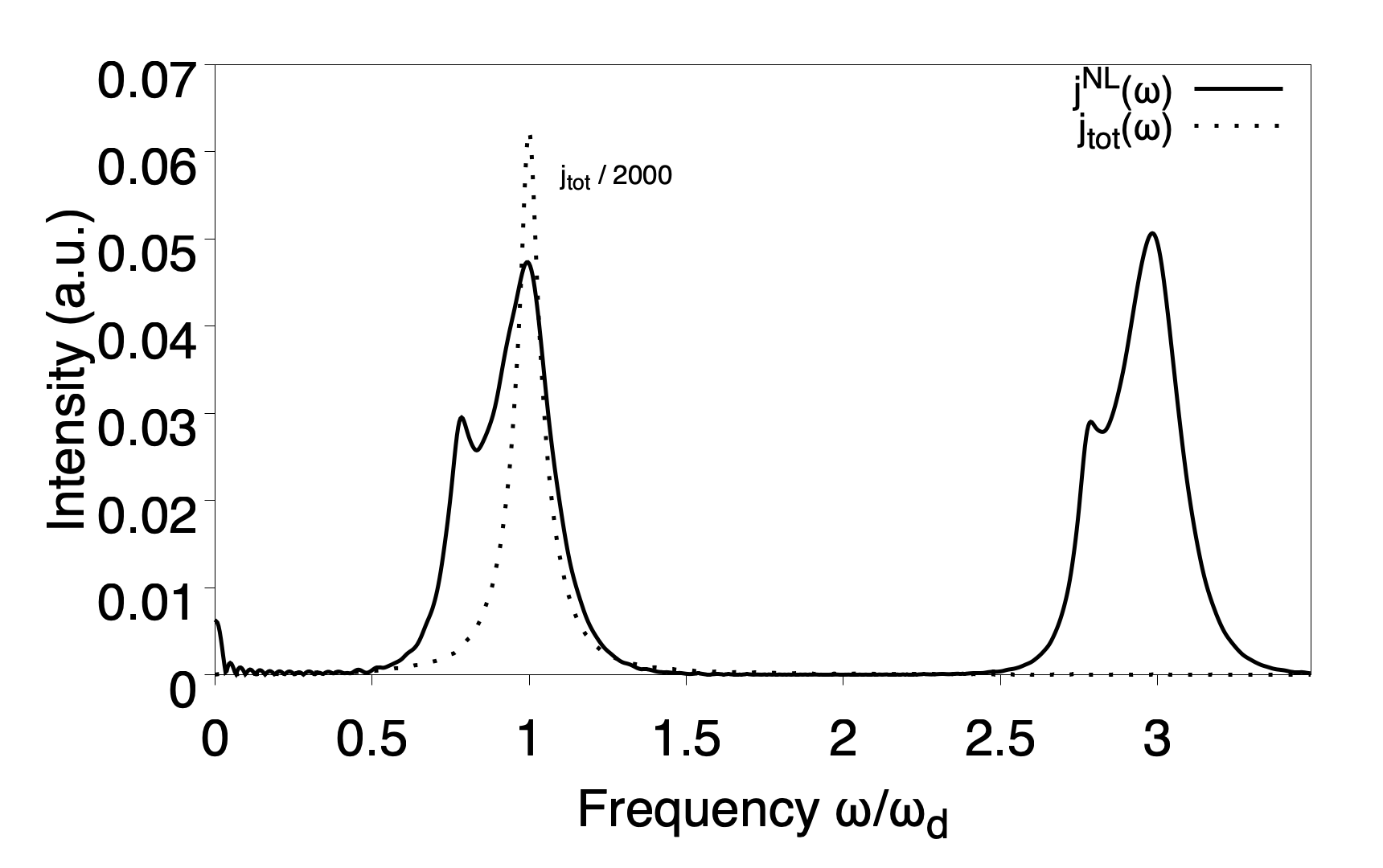}
\caption{\textbf{Driven high-harmonic generation.} Plot of the generated nonlinear (continuous line) and total (dashed line) current as a function of frequency, $j^{NL}(\omega)$ and $j_{tot} (\omega)$, respectively. This plot corresponds to a vertical cut in Fig. ~\ref{FFT2D} along $\omega$ for $\omega_{\Delta t} = 0$. The peaks at $\omega = \omega_d = 4.3$ THz and $\omega = 3 \omega_d = 12.9$ THz correspond to the fundamental and third harmonic, respectively. The smaller peak at $\omega = 12$ THz corresponds to the transient excitation of Higgs and quasiparticles with $\omega = \omega_d + 2 \Delta$, due to the asymmetric driving pulse.} \label{staticHHG}
\end{figure}

\begin{figure}[tb]
\centering
\includegraphics[width=8cm]{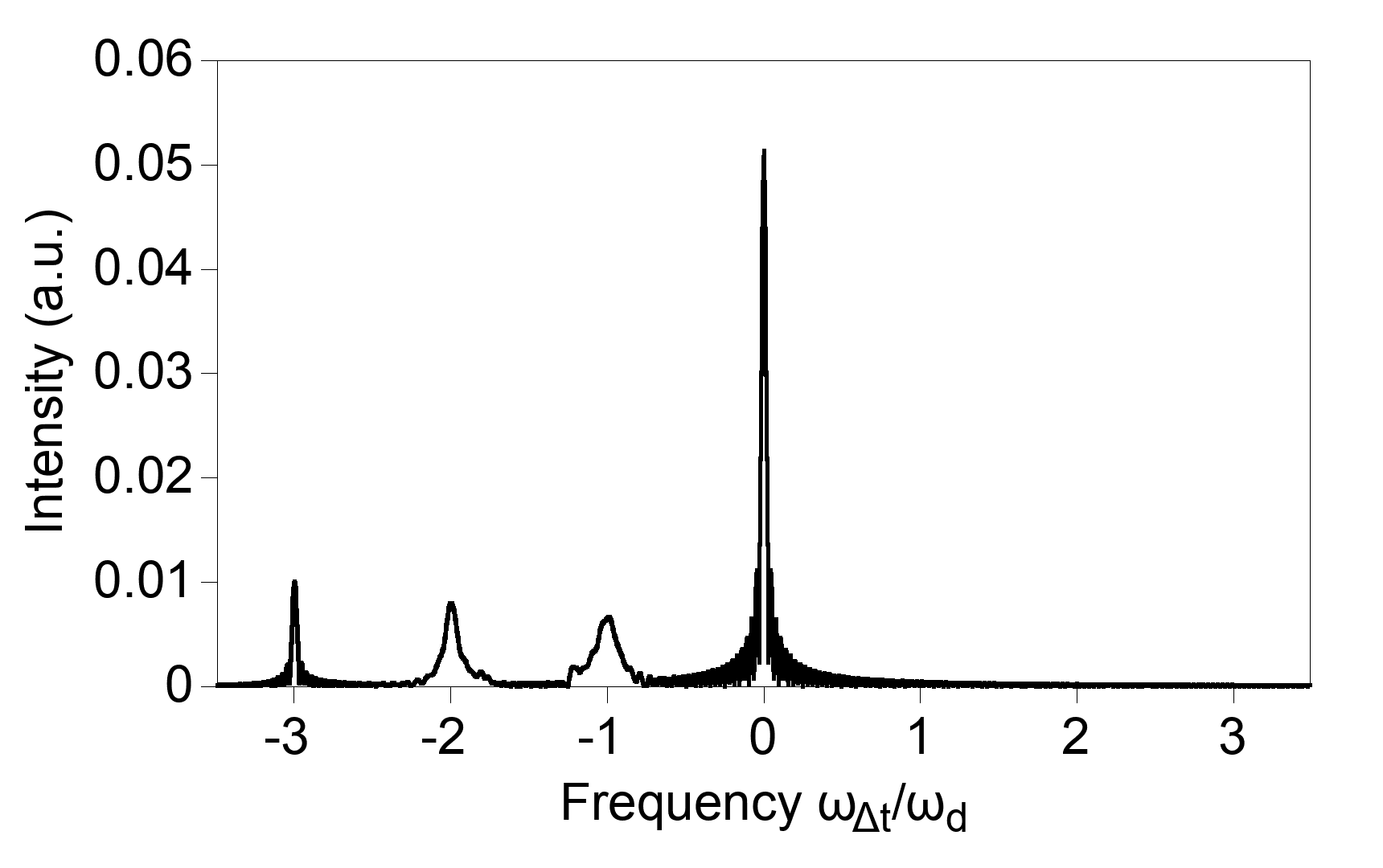}
\caption{\textbf{Transient modulation of higher harmonics.} Frequency-resolved spectral weight of the third harmonic current as a function of $\omega_{\Delta t}$, $j^{NL} (\omega_{\Delta t}, \omega = 3 \omega_d)$. This corresponds to a horizontal cut in the 2D Fourier-transform plot in Fig.~\ref{FFT2D} along $\omega_{\Delta t}$ at $\omega = 3 \ \omega_d$. It is possible to identify a transient modulation at frequencies $-\omega_d$, $-2 \omega_d$ and $- 3 \omega_d$, respectively.} \label{dynHM}
\end{figure}

\begin{figure}[tb]
\centering
\includegraphics[width=8cm]{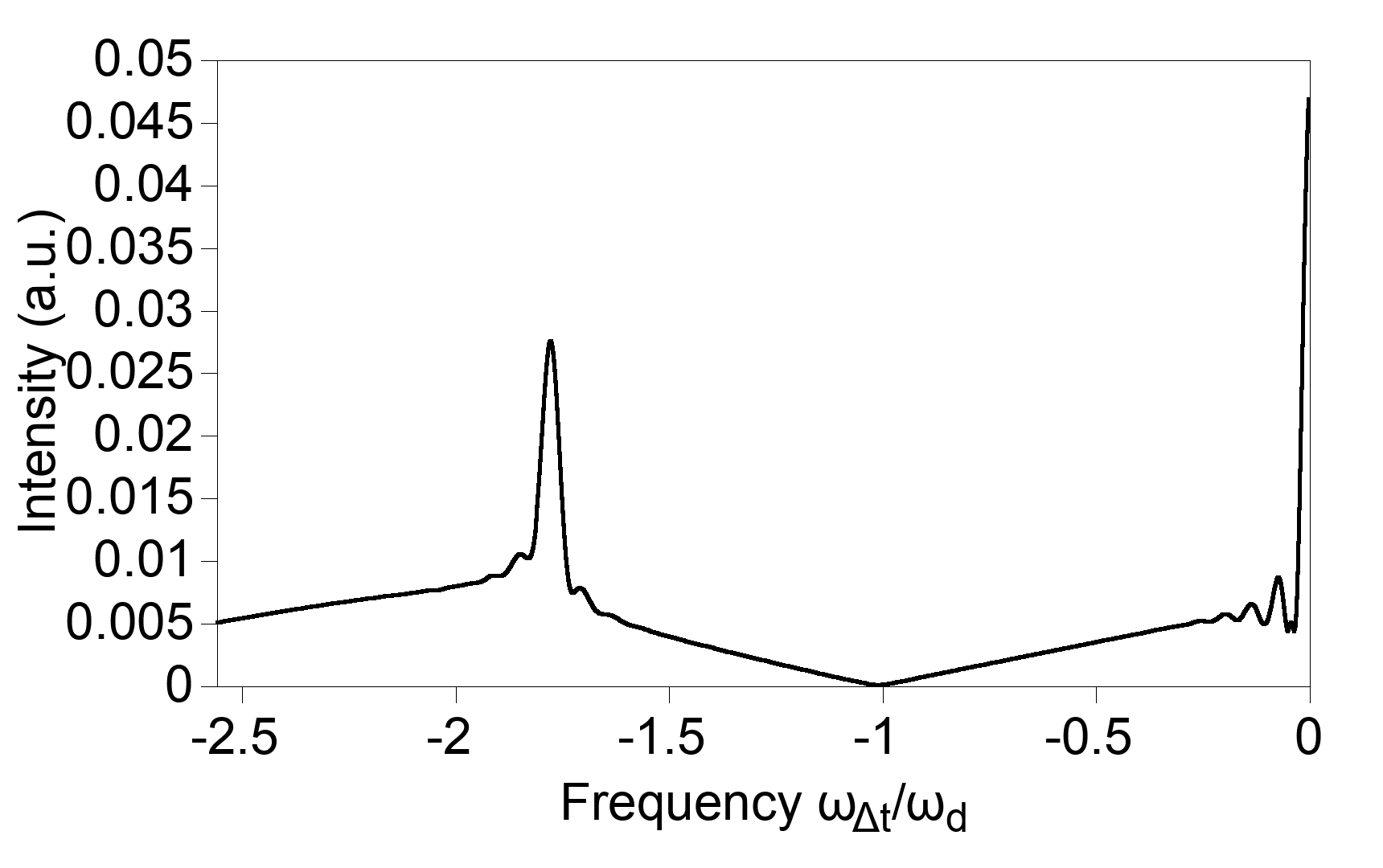}
\caption{\textbf{Transient excitation.} This plots corresponds to a diagonal cut in the 2D Fourier-transform plot in Fig.~\ref{FFT2D} along the line $ \omega = - \omega_{\Delta t} + \omega_d$. It is possible to identify a peak at a frequency $\omega_{\Delta t} = -7.7 \ \text{THz} = -1.8 \ \omega_d = -2 \Delta_0$.} \label{1Ddiag}
\end{figure}

To clearly distinguish the shape and position of the peaks observed in the frequency 2D plot, we performed one dimensional cuts of Fig.~\ref{FFT2D} along various lines. Fig.~\ref{staticHHG} shows the plot along the vertical line at $\omega_{\Delta t} = 0$. This corresponds to an equilibrium high-harmonic generation due to the driving pulse only. We observe a first harmonic peak at $\omega_d$ and a third harmonic signal at $3\omega_d$. Measurement of the temperature dependence of the THG peak would correspond to a usual THz THG experiment. 
The intensity of the fundamental and third harmonics (continuous
line) are of the same order, since only the third order is plotted. The total current has a dominant linear first harmonic response (dashed line).
In addition to the first and third harmonic, however, we notice the presence of
an additional shoulder peaks at a frequency $\omega = 2 \Delta + \omega_d$ and
at $\omega = 2 \Delta - \omega_d$. These are the result of the intrinsic Higgs
and quasiparticle resonance
at
$2 \Delta$ \cite{Collado}
in conjunction with a wave-mixing process with a driving photon of frequency
$\pm \omega_d$.
\\
In Fig.~\ref{dynHM} we show the prototypical case of a horizontal cut in
Fig.~\ref{FFT2D} along $\omega = 3 \omega_d$. The peak at $\omega_{\Delta
t}/\omega_d = 0$ is the equilibrium third harmonic as visible in
Fig.~\ref{staticHHG}, while the smaller peaks at $\omega_{\Delta t}/\omega_d =
-1,-2,-3$ stem from the modulation of the third harmonic due to the quench. Additional smaller peaks appear in Fig.~\ref{FFT1-2D} as a modulation in the delay time $\Delta t$ of the third harmonic, giving rise to the characteristic striped pattern. 
\\
Fig.~\ref{1Ddiag} corresponds to a diagonal line in the 2D plot of Fig.~\ref{FFT2D} passing through the point $\omega_{\Delta t} = 0, \omega = \omega_d$, and projected along the $\omega_{\Delta t}$ axis. The peak at $\omega_{\Delta t} = 0$ is the signal of the first harmonic. Of particular interest is the peak at $\omega_D = -2\Delta$, which is a direct consequence of the quasiparticle resonance at $\pm2\Delta$, represented by the process in Fig.~\ref{Diag}(d). 
Due to the wave mixing of the quench and the drive, we have here isolated the intrinsic superconducting response with the characteristic frequency of $2 \Delta$.\\
Moreover, the peaks in Fig.~\ref{FFT2D} along the diagonals placed at $\omega_{\Delta t} = - 2 \Delta + \omega_d$ are resulting from the process represented by the diagram in Fig.~\ref{Diag}(b), and they disappear when quench and drive have perpendicular polarization, since the corresponding interaction vertex vanishes (see Appendix \ref{Polarization}).

\section{Discussion} \label{discussion}
We can understand all features in the spectrum shown in Fig.~\ref{FFT2D}
by considering each of the diagrams in Fig.~\ref{Diag} that represent the
induced current expanded to third order in various combinations of powers of $A_d$ and $A_q$. 
The equilibrium THG signal proportional to $A_d^3$ has to be independent of $\Delta t$ and therefore falls onto the vertical line $\omega_{\Delta t}=0$ in the 2D spectrum $j(\omega,\omega_{\Delta t})$. This is represented by the diagram Fig.~\ref{Diag}$(c)$, where only the driving field acts on the condensate, and the current spectrum can be described as a function of real-time evolution as in Fig.~\ref{staticHHG}. Interestingly, we also notice that in addition to the aforementioned fundamental and third harmonic, a shoulder peak of the third harmonic at a frequency $\omega_d + 2 \Delta$ appears when the driving pulse is asymmetric and not Gaussian-shaped (see Appendix \ref{AppGauss} for the data with the symmetric envelope). This is the direct consequence of the effective quench induced by the driving, which launches free Higgs oscillations alongside the quasiparticle contribution and enhances the intensity of the nonlinear susceptibility at $\omega_H = 2 \Delta$ \cite{PhysRevB.92.064508,schwarz2020theory,wang2021transient}. In principle, a THG experiment 
for a single temperature would suffice to identify the collective mode resonance. However, this approach strictly relies on the condition $2\Delta \approx 2\omega_d$ and is specific to the asymmetric pulse shape \cite{wang2021transient} (See also Appendix~\ref{AppGauss}).
\\
The processes described by diagrams (a,b,d-f) in Fig.~\ref{Diag}, which involve at least one photon of the quench, 
are responsible for the signals along diagonal lines $\omega = \pm n
\omega_d - \omega_{\Delta t}$ with $n \in \{0,\pm1,\pm2\}$. The spectral window in which these lines can be observed are related to the bandwidth of the quench pulse. 
Here, we differentiate
between even and odd $n$. Odd diagonal lines show a peak at $\omega_{\Delta
t}=-2\Delta$ and even diagonals are peaked at $\omega_{\Delta t}=\omega_d -2\Delta$. This is expected from Eq.~\eqref{eqn:diagonal-lines-current} since the susceptibility $\chi_{\rho\rho}$ is 
peaked at $2\Delta$ for the modelled single-band superconductor. From Eq.~\ref{eqn:diagonal-lines-current}, we would additionally expect a peak at $\omega_{\Delta t}=-\omega_d -2\Delta$ for the line at $n=-2$. However, while the diagonal $n=-2$ line in principle is present, its spectral weight is negligibly small within the corresponding frequency range. Note that diagonal lines at $\pm n$ are related by frequency inversion $j(\omega,\omega_{\Delta t})=j(-\omega,-\omega_{\Delta t})$.

By inspecting the 2D spectrum in Fig.~\ref{FFT2D}, it is now straightforward to extract the
resonances of the nonlinear susceptibility $\chi_{\rho\rho}$. In our case, we observe the four peaks on the diagonal lines from which we extract the value $2\Delta$. If the superconducting condensate supports additional collective modes that nonlinearly couple in the electromagnetic response, their mode frequencies can be readily extracted as well.

\section{Conclusion and outlook} \label{conclusion}
In this work we proposed and analyzed from a new perspective a pump-probe spectroscopy setup on conventional clean superconductors with a combination of a single-cylce THz quench pulse and a multi-cycle driving THz probe field. We used a numerical approach based on the Anderson-pseudospin model to solve the equations of motion and to calculate the generated nonlinear current. In addition, we investigated the nonlinear optical processes by means of a diagrammatic approach to interpret and explain the obtained results. 
\\
In particular, we showed that, in addition to the usual third harmonic
generation measured in driving experiments, new features are obtained in a two-dimensional spectrum of the nonlinear generated current. These features are manifest 
as diagonal lines in 2D frequency space of the time and pump-probe delay and allow for a direct extraction of resonances in the nonlinear susceptibility.
The susceptibility encodes the intrinsic superconducting response of the quasiparticles and resonances of transient excitation of the Higgs mode. \\

The advantage of a two-dimensional analysis of quench-drive
spectroscopy is manifest in the possibility to scan a wider frequency spectrum
at once with fixed parameters of quench and drive pulses, by scanning the
quench-drive delay time. In addition, with the present setup the quench pulse
allows to push the system out of equilibrium, quenching and shrinking the superconducting gap, allowing the driving pulse to probe different states of the superconductor, resulting in different peak profiles and positions in the 2D frequency spectra.

It is also interesting to examine the possibility to extend the quench-drive spectroscopy framework to the case of cuprates, which exhibit a different symmetry of the order parameter in momentum space, and pre-formed phase-incoherent Cooper pairs \cite{D2FD00010E}, which can reveal more information on the competing orders and their symmetries.

All in all, we believe that this work can pave the way towards coherent time-dependent multi-dimensional spectroscopy on superconductors in the THz regime. A full two-dimensional pump-pump-probe spectroscopy with coherent pulses will be the focus of a future work. Its possibilities range from coherent control of superconductors to the study of competing orders, such as superconductivity, charge-density wave, and bi-plasmon among others \cite{Cundiff2013,Giannetti2016}.
Other systems where the Higgs response is known to be enhanced, such as cuprates, could be interesting to investigate with this spectroscopic approach, to efficiently study the transient non-equilibrium response of quasiparticles and the Higgs mode and to unveil the features of their rich phase diagram.

\acknowledgements
Fruitful discussions with L. Benfatto, P. M. Bonetti, S. Kaiser, M.-J. Kim, L. Schwarz are thankfully acknowledged. We thank the Max Planck-UBC-UTokyo Center for Quantum Materials for fruitful collaborations and financial support. R.H. acknowledges the Joint-PhD program of the University of British Columbia and the University of Stuttgart. 

\appendix
\onecolumngrid
\section*{APPENDIX}

\section{Pseudospin model} \label{apdx:pseudospin}
In this Appendix we will derive the equations of motion for the order parameter
within the pseudospin formalism. We note that all results
shown in the paper are computed within a fully numerical approach that does not
rely on any numerical approximation. 

We first write the BCS Hamiltonian using the pseudospin formalism as
\begin{align}
\hat{H} = \sum_\vk \mathbf{b}_\vk \hat{\mathbf{\sigma}}_\vk \,,
\end{align}
with the pseudospin
\begin{align}
\hat{\mathbf{\sigma}}_\vk = \frac{1}{2} \hat{\Psi}_\vk^\dagger \mathbf{\tau} \hat{\Psi}_\vk \,,
\end{align}
$\mathbf{\tau} = (\tau_1, \tau_2, \tau_3)$ being the vector of the Pauli matrices $\tau_{i=1,2,3}$, and the pseudo-magnetic field
\begin{align}
\mathbf{b}_\vk = (- \Delta' f_\vk, - \Delta'' f_\vk,  \eps) \,,
\end{align}
provided the gap equation
\begin{align}
\Delta_\vk = \Delta f_\vk = V f_\vk \ \sum_{\vk'} f_{\vk'} \langle \hat{c}_{-\vk', \downarrow} \hat{c}_{\vk', \uparrow} \rangle \,.
\end{align}
Here $\hat{c}$ is the electronic annihilation operator, $f_\vk$ is the superconducting form factor, $\eps$ is the electronic band dispersion.
In the presence of an external field represented by the vector potential $\mathbf{A}(t)$, we get
\begin{align}
\mathbf{\sigma}_\vk (t) = \mathbf{\sigma}_\vk (0) + \delta \mathbf{\sigma}_\vk (t) \,, 
\end{align}
with $\delta \mathbf{\sigma}_\vk (t) = (x_\vk (t), y_\vk (t),  z_\vk (t))$, and 
\begin{align}
\mathbf{b}_\vk (t) = (- \Delta'(t) f_\vk,  - \Delta''(t) f_\vk,  \varepsilon_{\vk - e \mathbf{A}(t)} + \varepsilon_{\vk + e \mathbf{A}(t)}) \,.
\end{align}
The equation of motion for the pseudo-magnetic field can be written in Bloch form, $\partial_t \mathbf{\sigma}_{\vk} = 2 \textbf{b}_{\textbf{k}} \times \mathbf{\sigma}_{\textbf{k}}$. We now write the three components of the Bloch equation by applying to the time-dependent pseudospin the ansatz described above:
\begin{eqnarray}
    \left\{
    \begin{array}{ll}
    \begin{split}
        \sigma_x (t) &=  \sigma_x^{eq} + x(t) \\    
        \sigma_y (t) &=  y(t) \\    
        \sigma_z (t) &=  \sigma_z^{eq} + z(t) \,,
    \end{split}
    \end{array}
    \right.
\end{eqnarray}
with $\sigma_y^{eq} = 0$ under the assumption of real order parameter, $\sigma_x^{eq} = \Delta / (2 E_{\textbf{k}})$, $\sigma_z^{eq} = - \varepsilon_{\textbf{k}} / (2 E_{\textbf{k}})$, where the quasiparticle energy dispersion is $E_{\textbf{k}} = \sqrt{\varepsilon_{\textbf{k}}^2 + \Delta^2 f_\vk^2}$. \\
Thus, the Bloch equations result in
\begin{eqnarray}
    \left\{
    \begin{array}{ll}
    \begin{split}
        \partial_t x(t) &= - (\varepsilon_{\vk - e \mathbf{A}} + \varepsilon_{\vk + e \mathbf{A}}) y(t) - \dfrac{f_\vk}{\E} \eps \delta \Delta'' (t) + 2 \delta \Delta'' (t) f_\vk z(t) \,, \\
        \partial_t y(t) &= 2 \varepsilon_{\textbf{k}} x(t) + 2 ( \Delta + \delta \Delta' (t) ) f_\vk z(t) - \ \delta \Delta' f_\vk \dfrac{\eps}{\E} + \dfrac{\Delta f_\vk}{2 \E} (\varepsilon_{\vk - e \mathbf{A}} + \varepsilon_{\vk + e \mathbf{A}} - 2 \eps) \,, \\
        \partial_t z(t) &= -2 \ \Delta f_\vk \ y(t) - \dfrac{\Delta f_\vk^2}{\E} \delta \Delta'' (t) - 2 \delta \Delta'' (t) f_\vk x(t) \,.
    \end{split}
    \end{array}
    \right. \label{SelfC}
\end{eqnarray}
with $\mathbf{A}(t) = \mathbf{A}_{q} (t) + \mathbf{A}_{d} (t)$, $\delta \Delta (t) = \delta \Delta'(t) + i \ \delta \Delta''(t)$ being the time-dependent variation of the order parameter induced by the external field, such that $\Delta (t) = \Delta + \delta \Delta (t)$. We introduce now the quench-drive delay time $\Delta t = t_d - t_q$, which measures the time distance of the peaks of the two pulses, and we put $t_d=0$, so that we can rewrite $\mathbf{A}(t) = \mathbf{\overline{A}}_{q} (t + \Delta t) + \mathbf{\overline{A}}_{d} (\overline{t})$. 

\begin{figure}[htb]
\centering
\includegraphics[width=14cm]{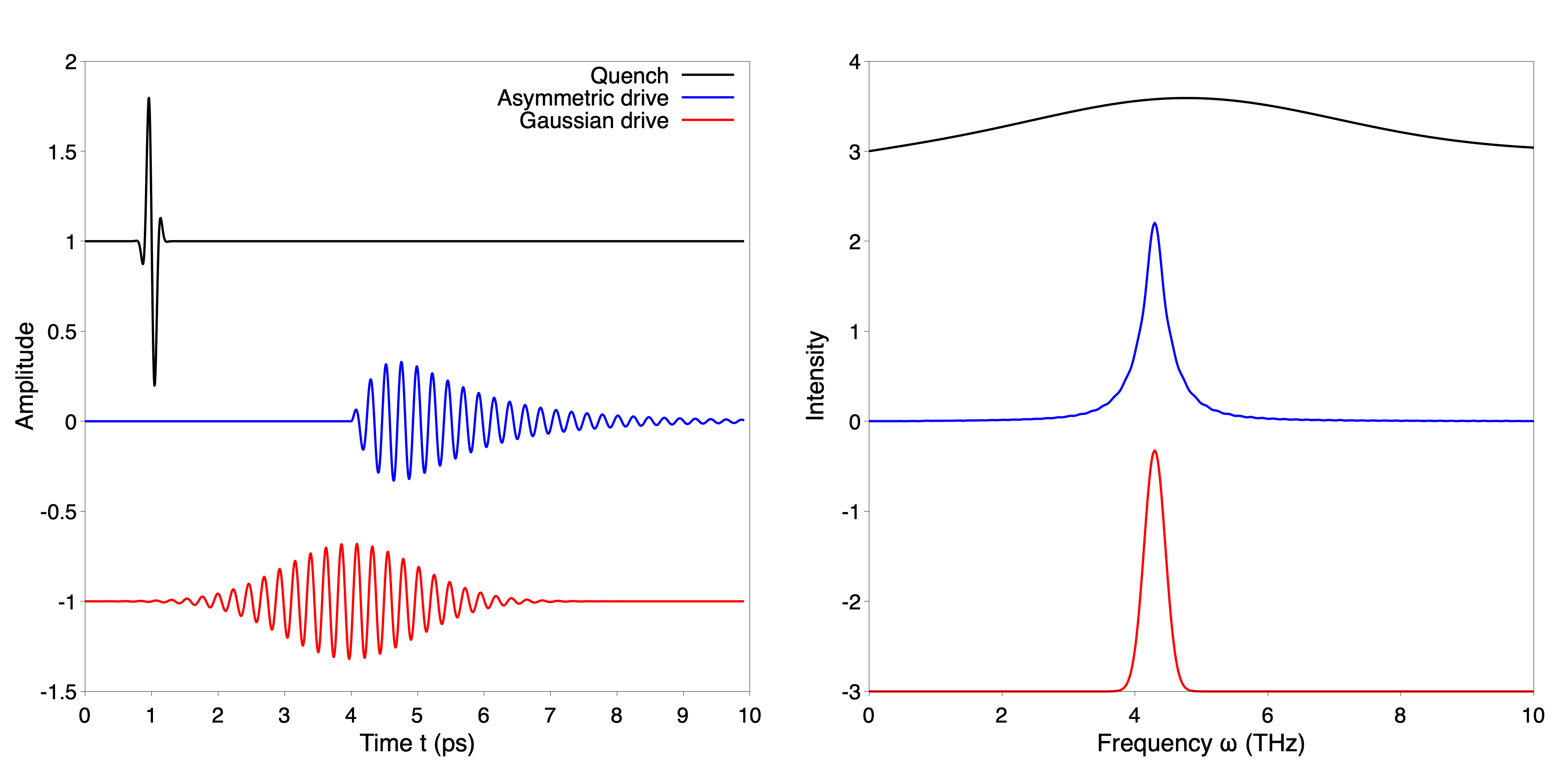}
\caption{\textbf{Quench and drive pulses.} Plot of the amplitude of the vector potential $A$ corresponding to the quench (top left, frequency $\omega_q = 4.77$ THz) and drive pulse, respectively, used in the calculations: the asymmetric drive pulse (center left) and Gaussian shaped drive pulse (bottom left) have a frequency $\omega_d = 4.3$ THz. On the right their Fourier spectrum in frequency is shown: the asymmetric drive has a sharp peak in frequency with a slower decay, while the Gaussian-shaped drive is narrower in frequency.} 
\label{pulses}
\end{figure}

\section{Quench and drive pulses} \label{pulsesDetails}
As mentioned in the main text, in the quench-drive setup we adopted three
different kinds of pulses: a few-cycles quench, and two different kind of
drives, either asymmetric or with a Gaussian envelope profile [Fig.
(\ref{pulses})], all of them polarized in $x$ direction (due to the symmetry of the superconductor order parameter, this assumption can be made without loss of generality).\\
The quench pulse is characterized by a very few cycles and its vector potential can be written as
\begin{align}
A_q(t) = A^0_q \ e^{-(t-t_q)^2/\tau_q^2} \cos{(\omega_q (t-t_q))} \,,
\end{align}
with $\tau_q = 0.1$ ps and the quench central frequency $\omega_q = 4.77$ THz.\\
The asymmetric driving pulse is defined by the expression
\begin{eqnarray}
A_d(t) = 
    \left\{
    \begin{array}{l}
    \begin{split}
        & A^0_d \dfrac{\sin{(\omega_d (t-t_d))}}{1+(t-t_d)^2} (t-t_d) \ e^{-(t-t_d)/\tau_d} \quad \,, \quad \text{for} \ t \geq t_d \,, \\  
        & 0 \quad \,, \quad \text{for} \ t < t_d \,,
    \end{split}
    \end{array}
    \right.
\end{eqnarray}
with $\tau_d = 2$ ps and the driving central frequency $\omega_d = 4.3$ THz.
The driving pulse reaches its maximum intensity after $\approx$ 1 ps, and then
slowly decays within $\approx$ 5 ps. Therefore, the initial part of the drive
acts as an effective quench on the superconductor, while the its decay drives the condensate and its collective modes. \\
Similar to the quench pulse, the time-symmetric drive with Gaussian envelope is 
\begin{align}
A_d(t) = A^0_d e^{-(t-t_0)^2/\tau_d^2} \cos{(\omega_d (t-t_0))} \,.
\end{align}

\newpage 
\section{Cross polarization of quench and drive pulses} \label{Polarization}
\begin{figure}[ht]
\includegraphics[width=12cm]{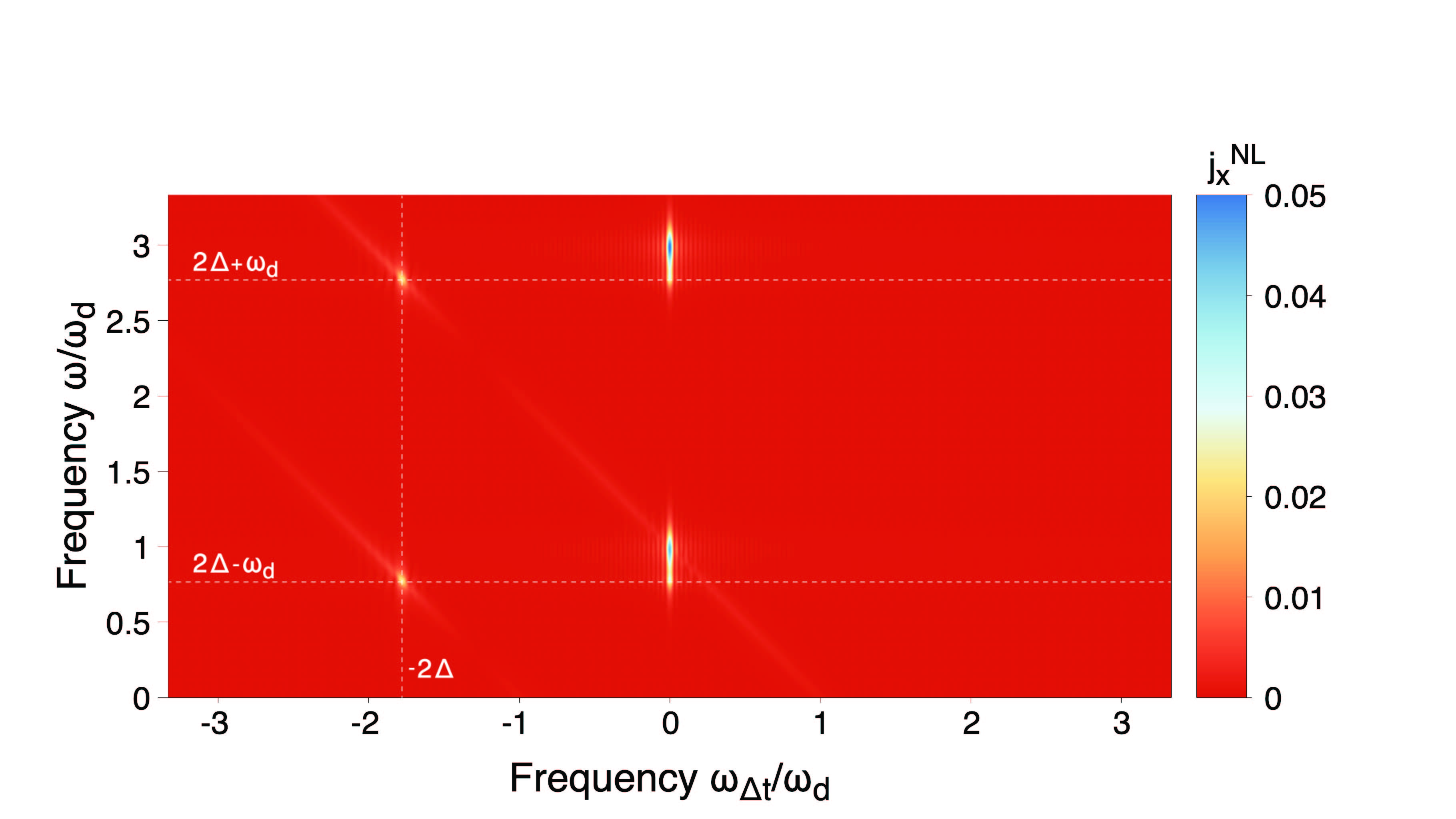}
\includegraphics[width=12cm]{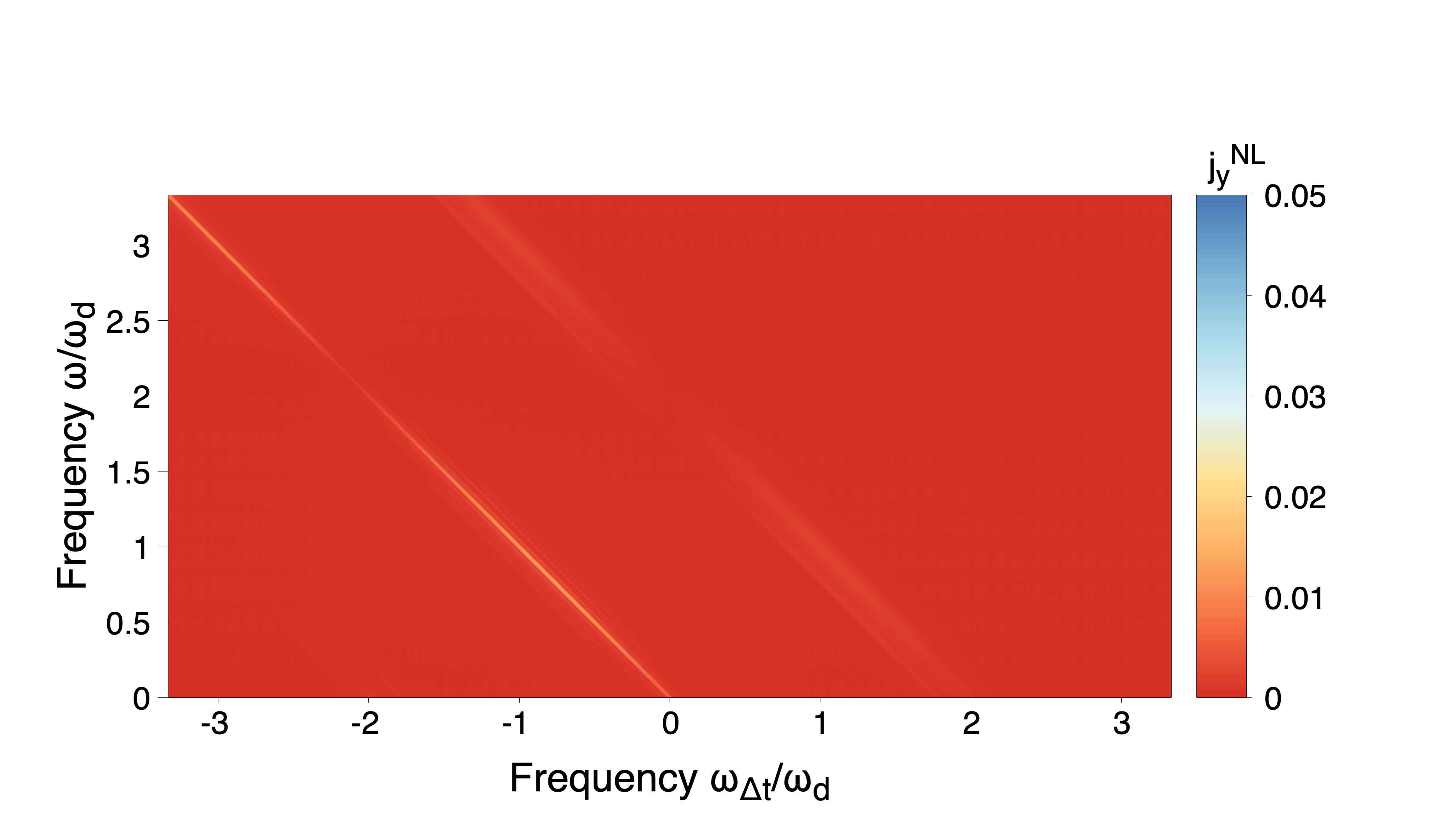}
\caption{\textbf{Nonlinear current spectra for cross-polarized pulses.} 2D Fourier transform of $x-$ (top) and $y-$component (bottom) of the nonlinear current response for cross-polarized quench (linearly polarized along the $x$ axis) and drive (linearly polarized along the $y$ axis).} \label{FFT2Dcross}
\end{figure}

\begin{figure}[ht]
\includegraphics[width=8cm]{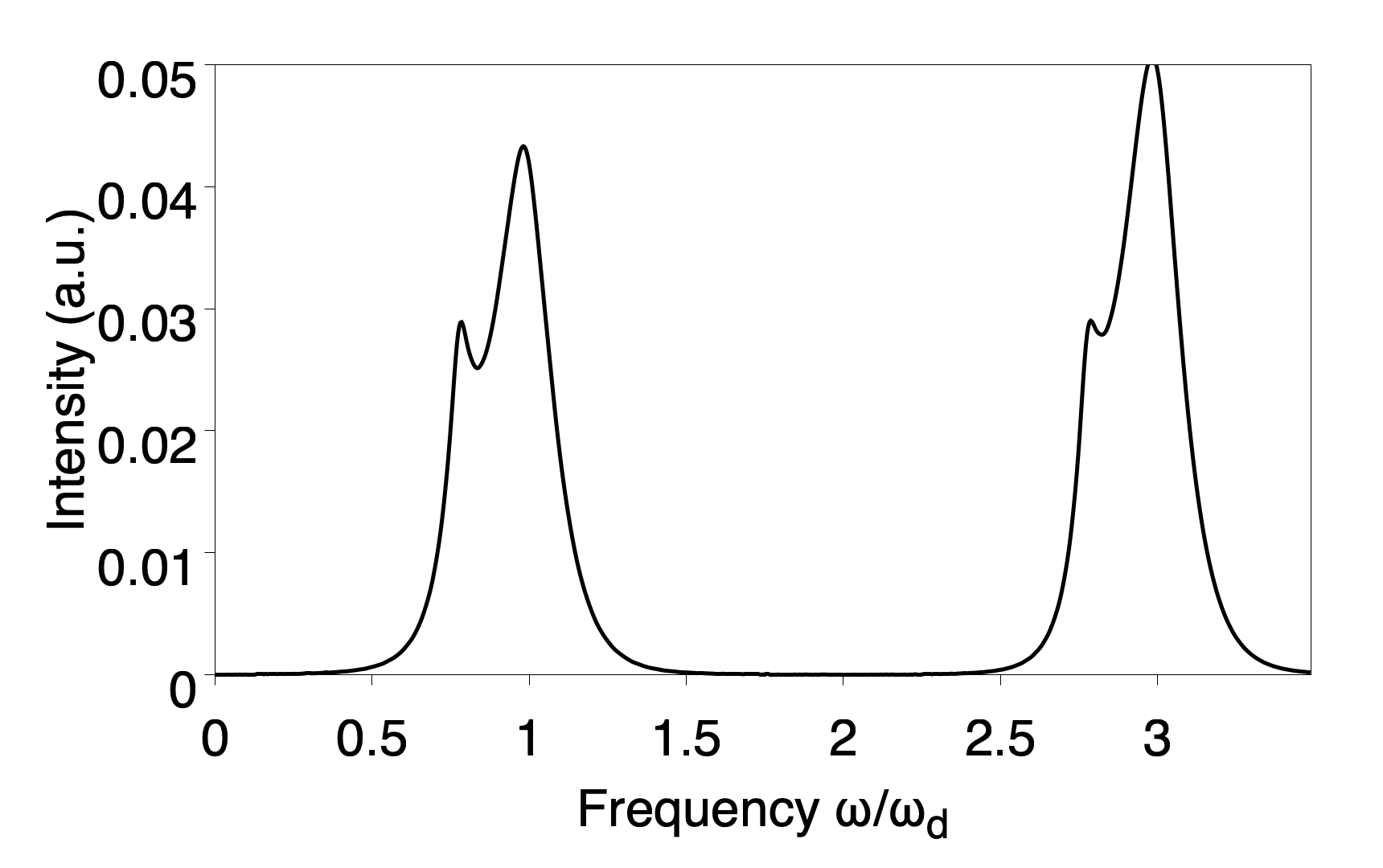}
\includegraphics[width=8cm]{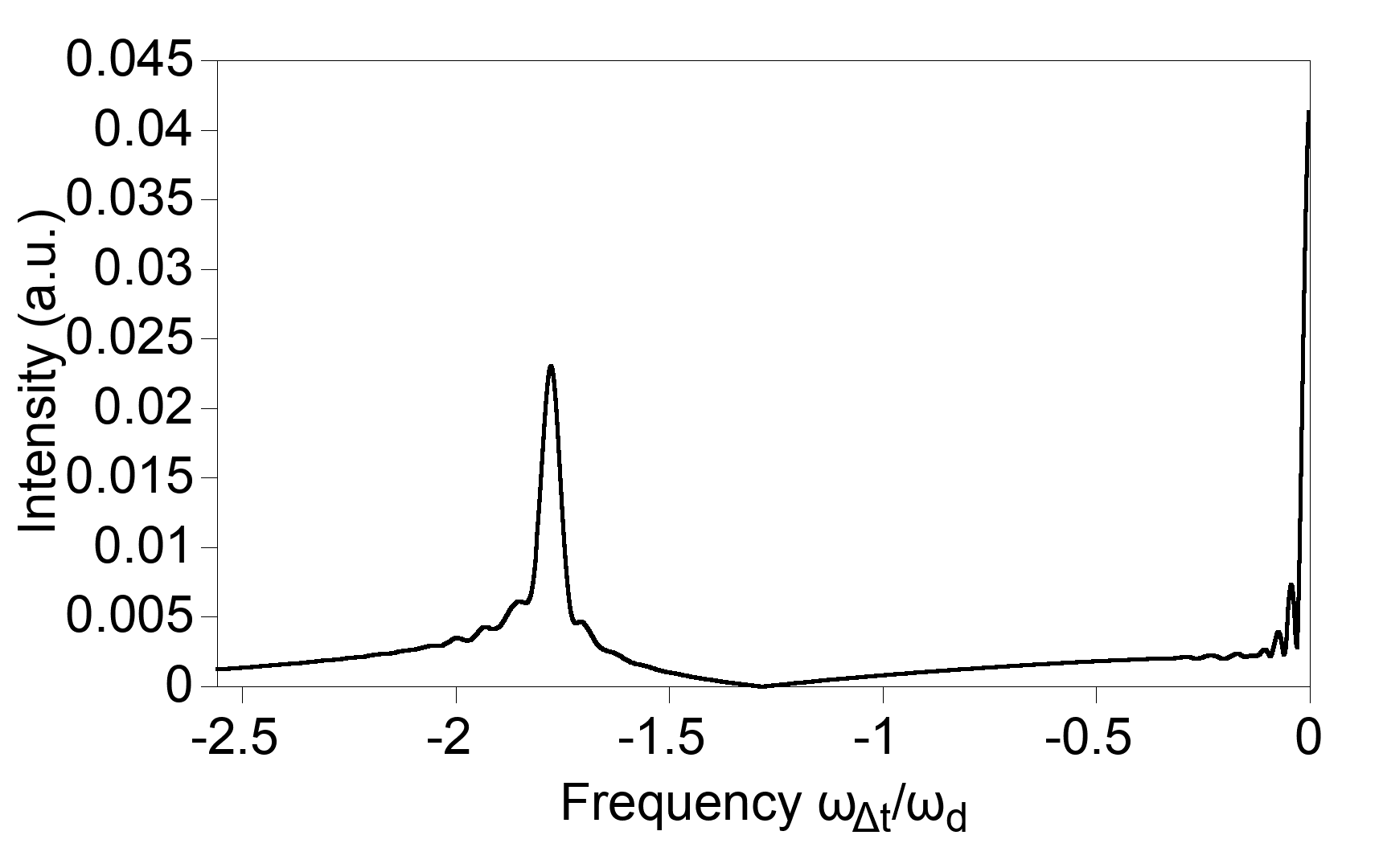}
\caption{\textbf{Nonlinear current along the $x$ axis: traces from the 2D
Fourier transform.} Left: Nonlinear current as a function of the frequency
$\omega$, obtained with a trace along the vertical axis from the left plot in
Fig.\ref{FFT2Dcross}. Right: Nonlinear current as a function of $\omega_{\Delta
t}$ with the constraint $\omega = \omega_d - \omega_{\Delta t}$, obtained from a
diagonal cut passing by the fundamental harmonic of the left plot in Fig.~\ref{FFT2Dcross}. } \label{1DcutX}
\end{figure}

So far, we have analyzed the setup with parallel
		linearly-polarized quench and drive pulses along the $x$ axis.
For $s$-wave BCS superconductors, the direction of the polarization of the pulses is arbitrary and does not affect the results, as long as the polarization is kept parallel. \\
In this section, we study the case of non-parallel quench and drive pulses. The
drive is fixed along the $x$ axis, the quench is linearly polarized along the
$y$ axis. We calculate the output current along the two orthogonal axes $x$ and
$y$. The corresponding spectra are plotted in Fig.~\ref{FFT2Dcross} as a function
of $\omega$ and $\omega_{\Delta t}$. The fundamental and third harmonic signals
are still present in the Fourier transform of the current along $x$, together
with the side-peaks at $2 \Delta \pm \omega_d$. Along the diagonals, the peaks
at $\omega_{\Delta t} = - 2 \Delta_0 + \omega_d$ are absent, as shown in
Fig.~\ref{1DcutX} (in contrast to the parallel polarization, as shown in
Fig.~\ref{1Ddiag}), while the ones located at $\omega_{\Delta t} = - 2 \Delta_0
$ are still present. This is due to the fact that the quench and the drive
cannot interact in the same vertex as in Fig.~\ref{Diag}(b) since they act on
perpendicular directions, while the contribution of two quench pulses, which
provides a peak at a frequency of $-2 \Delta_0$ (as in Fig.~\ref{Diag}(d)), can still take place. \\
 
\newpage 
\section{Symmetric Gaussian-shaped driving pulse} \label{AppGauss}
\begin{figure}[htb]
\centering
\includegraphics[width=14cm]{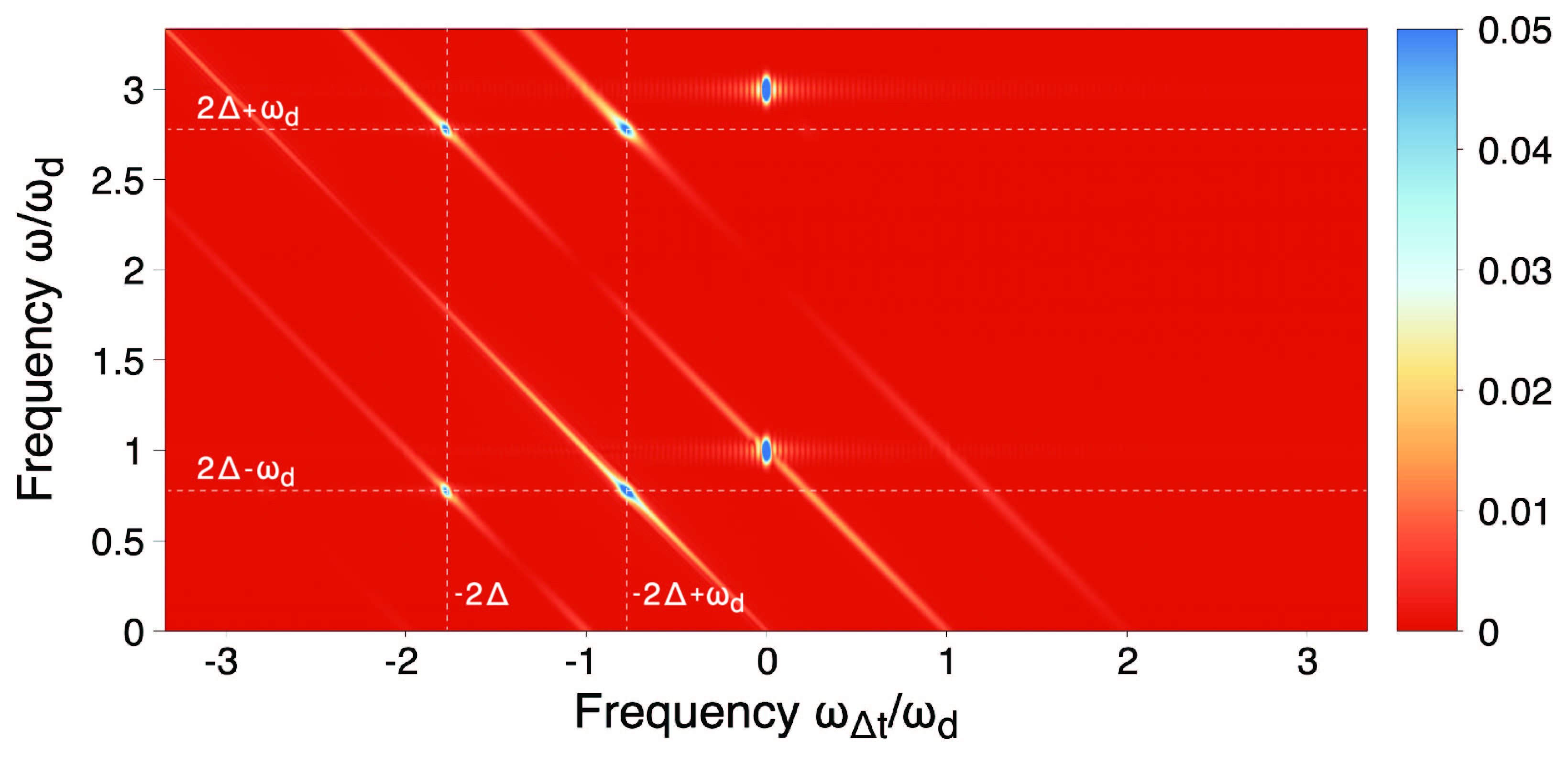}
\caption{\textbf{Two-dimensional Fourier-transformed plot of the nonlinear current.} 2D plot of the generated nonlinear output current intensity $j^{NL}(\omega, \omega_{\Delta t})$ as a function of the real frequency $\omega$ and the quench-drive delay frequency $\omega_{\Delta t}$. It corresponds to the two-dimensional Fourier transform of the data in Fig.~\ref{2Dtt}. The vertical response at $\omega_{\Delta t}=0$ corresponds to the quench-free superconducting signal, namely the high-harmonic generation due to the driving field. The diagonal lines, instead, represent the transient modulation of the higher-harmonics due to the quench-drive wave mixing. The symmetric Gaussian shaped driving pulse was used here.} \label{FFT2Dg}
\end{figure}

\begin{figure}[htb]
\centering
\includegraphics[width=8cm]{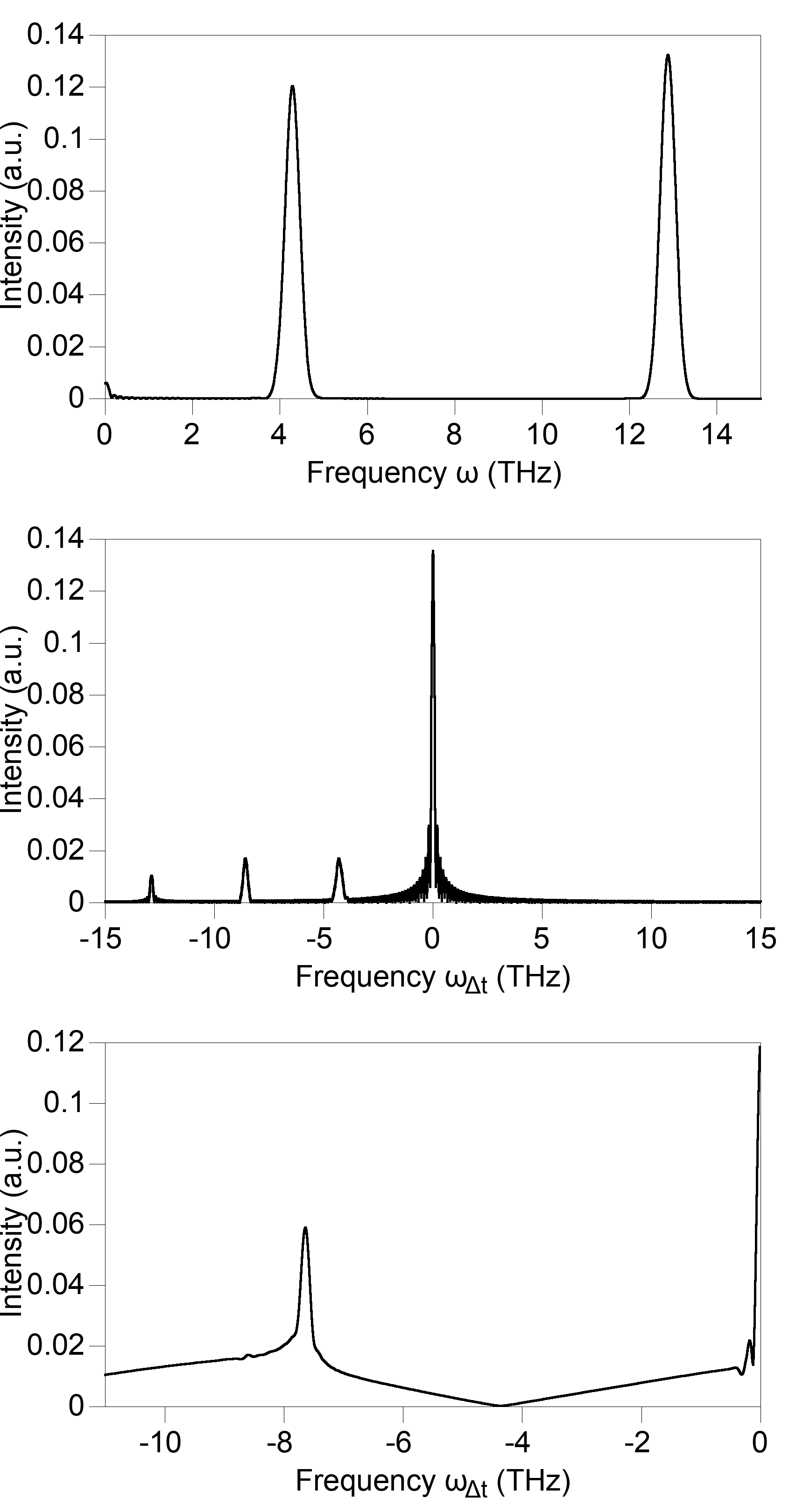}
\caption{\textbf{High-harmonic generation and transient modulation with Gaussian drive.} Top: nonlinear current $j^{NL} (\omega)$ at $\omega_{\Delta t} = 0$. In contrast to Fig.~\ref{staticHHG}, no peak at $\omega = 2 \Delta + \omega_d$ is present in this case. Center: nonlinear current modulation as a function of the quench-drive frequency $\omega_{\Delta t}$, $j^{NL} (\omega_{\Delta t}, \omega = 3 \omega_d)$. This corresponds to Fig.~\ref{dynHM} using the Gaussian envelope for the drive pulse. Bottom: $j^{NL} (\omega_{\Delta t})$ obtained with the condition $\omega = - \omega_{\Delta t} + \omega_d$, equivalent to Fig.~\ref{1Ddiag}.} \label{1Dmulti}
\end{figure}

Here, we repeated the same calculations of the nonlinear current generation in
the quench-drive spectroscopy setup as in the main text, but using a symmetric
Gaussian envelope for the drive pulse. The results are shown in
Figs.~\ref{FFT2Dg}-\ref{1Dmulti} and correspond to those in
Figs.~\ref{FFT2D}-\ref{1Ddiag} in the main text, where the asymmetric drive was
used. In the top panel of Fig.~\ref{1Dmulti}, the equilibrium
high-harmonic generation does not include the shoulder peak at $\omega = 2 \Delta +
\omega_d$, $\omega_{\Delta t}=0$, since it was generated by the initial effective quench of the asymmetric driving field. However, all other features of high-harmonic modulation and transient excitation at $\omega_{\Delta t} = 2 \Delta$ are still present, since they originate from the wave mixing of the quench and the drive pulses, independently of their shape.

\twocolumngrid

\end{document}